\documentclass[journal,transmag]{IEEEtran}

\usepackage{color}
\usepackage{float}
\usepackage{graphicx}
\usepackage{subcaption}
\usepackage{rotating}
\usepackage{amsmath}
\usepackage{amsfonts}
\usepackage{todonotes} 
\usepackage{hyperref}

\input{Qcircuit.tex }

\definecolor{red}{rgb}{1,0.,0}
\newcommand{\inp}[2]{\left< #1|#2 \right>} 

\title{Quantum Computing Circuits and Devices}
\date{Started 19 NOV 2017}

\author{\IEEEauthorblockN{Travis~S.~Humble\IEEEauthorrefmark{1},
        Himanshu Thapliyal\IEEEauthorrefmark{2},
        Edgard~Mu\~{n}oz-Coreas\IEEEauthorrefmark{2}, \\
        Fahd~A.~Mohiyaddin\IEEEauthorrefmark{1},
        Ryan S.~Bennink\IEEEauthorrefmark{1}}\\%
\IEEEauthorblockA{\IEEEauthorrefmark{1}Quantum Computing Institute, Oak Ridge National Laboratory, Oak Ridge, Tennessee USA \\
		\IEEEauthorrefmark{2}Department of Electrical \& Computer Engineering, University of Kentucky, Lexington, Kentucky USA}
\thanks{This manuscript has been authored by UT-Battelle, LLC, under Contract No. DE-AC0500OR22725 with the U.S. Department of Energy. The United States Government retains and the publisher, by accepting the article for publication, acknowledges that the United States Government retains a non-exclusive, paid-up, irrevocable, world-wide license to publish or reproduce the published form of this manuscript, or allow others to do so, for the United States Government purposes. The Department of Energy will provide public access to these results of federally sponsored research in accordance with the DOE Public Access Plan (http://energy.gov/downloads/doe-public-access-plan).}
}

\begin{document}

\maketitle

\begin{abstract}
The development of quantum computing technologies builds on the unique features of quantum physics while borrowing familiar principles from the design of conventional  devices. We introduce the fundamental concepts required for designing and operating quantum computing devices by reviewing state of the art efforts to fabricate and demonstrate quantum gates and qubits. We summarize the near-term challenges for devices based on semiconducting, superconducting, and trapped ion technologies with an emphasis on design tools as well as methods of verification and validation. We then discuss the generation and synthesis of quantum circuits for higher-order logic that can be carried out using quantum computing devices. 
\end{abstract}


\section{Introduction}
\label{sec:introductions}
%
Quantum computing promises new capabilities for processing information and performing computationally hard tasks.  This includes significant algorithmic advances for solving hard problems in computing \cite{Montanaro2016}, sensing \cite{Degen2017}, and communication \cite{Krenn2016}. The breakthrough examples of Shor's algorithm for factoring numbers and Grover's algorithms for unstructured search have fueled a series of more recent advances in computational chemistry, nuclear physics, and optimization research among many others. However, realizing the algorithmic advantages of quantum computing requires hardware devices capable of encoding quantum information, performing quantum logic, and carrying out sequences of complex calculations based on quantum mechanics \cite{Nielsen2010}.  For more than 35 years, there has been a broad array of experimental efforts to build quantum computing devices to demonstrate these new ideas. Multiple state-of-the-art engineering efforts have now fabricated functioning quantum processing units (QPUs) capable of carrying out small-scale demonstrations of quantum computing. The QPUs developed by commercial vendors such as IBM, Google, D-Wave, Rigetti, and IonQ are among a growing list of devices that have demonstrated the fundamental elements required for quantum computing \cite{Linke2017}. This progress in prototype QPUs has opened up new discussions about how to best utilize these nascent devices \cite{Britt2017,Britt2017QA}.
\par
Quantum computing poses several new challenges to the concepts of design and testing that are unfamiliar to conventional CMOS-based computing devices. For example, a striking fundamental challenge is the inability to interrogate the instantaneous quantum state of these new devices. Such interrogations may be impractically complex within the context of conventional computing, but they are physically impossible within the context of quantum computing due to the no-cloning principles. This physical distinction fundamentally changes how QPUs are designed and their operation tested relative to past practice. This tutorial provides an overview of the principles of operation behind quantum computing devices as well as a summary of the state of the art in QPU. The continuing development of quantum computing will require expertise form the conventional design and testing community to ensure the integration of these non-traditional devices into existing design workflows and testing infrastructure. There is a wide variety of technologies under consideration for device development, and this tutorial focuses on the current workflows surrounding quantum devices fabricated in semiconducting, superconducting, and trapped ion technologies. We also discuss the design of logical circuits that quantum devices must execute to perform computational work.
\par
While the tutorial captures many of the introductory topics needed to understand the design and testing of quantum devices, several more advanced topics have been omitted due to space constraints. Foremost is the broader theory of quantum computation, which has developed rapidly from early models of quantum Turing machines to a number of different but equally powerful computational models. In addition, we have largely omitted the the sophisticated techniques employed to mitigate the occurrence of errors in quantum devices. Quantum error correction is an important aspect of long-term and large-scale quantum computing, which uses redundancy to overcome the loss in information from noisy environments. Finally, our review of quantum computing technologies is intentionally narrowed to three of the leading candidates for large-scale quantum computing. However, there is a great diversity of experimental quantum physical systems that can be used for encoding and processing quantum information.  
\par
The tutorial is organized as follows: Sec.~\ref{sec:principles} provides an introduction to the principles of quantum information and quantum computing; Sec.~\ref{sec:devices} provides an overview of several quantum computing devices and their use in developing quantum processing units; Sec.~\ref{sec:vv} discusses concerns for the verification and validation of these devices;  Sec.~\ref{sec:circuits} provides a similar presentation for the specification and design of quantum circuits; and Sec.~\ref{sec:conclusion} offers a summary of future developments.

\section{Principles of Quantum Computing}
\label{sec:principles}
%
The principles of quantum computing derive from quantum mechanics, a theoretical framework that has accurately modeled the microscopic world for more than 100 years. Quantum computing draws its breakthroughs in computational capabilities from the many unconventional features inherent to quantum mechanics, and we provide a brief overview of these features while others offer more exhaustive explanations \cite{Nielsen2010}.
\par
In quantum mechanics, all knowable information about a physical system is represented by a \emph{quantum state}. The quantum state is defined as vector within a Hilbert space, which is a complex-valued vector space supporting an inner product. By convention, the quantum state with label $\Psi$ is expressed using the `ket' notation as $\ket{\Psi}$, while the dual vector is expressed as the `bra' $\bra{\Psi}$. The inner product between these two vectors is $\inp{\Psi}{\Psi}$ and normalized to one. An orthonormal basis for an $N$-dimensional Hilbert space satisfies $\inp{i}{j} = \delta_{i,j}$, and an arbitrary quantum state may be represented within a complete basis as
\begin{equation}
\label{eq:psi}
\ket{\Psi} = \sum_{j=0}^{N-1}{c_{j} \ket{j}},
\end{equation}
where $c_{j} = \inp{j}{\Psi}$ is the corresponding coefficient. Within a chosen basis, the coefficients of the quantum state are interpreted as probability amplitudes such that the squared magnitude of this amplitude yields the probability to lie along the chosen basis, i.e., $p_j = |c_{j}|^2$. The mathematical theory of quantum mechanics is exceedingly rich and draws from aspects of linear algebra, probability, and complex analysis. Additional details on these aspects points are found, e.g., in Ref.~\cite{Sakurai1995}.
\par
The fundamental equation of motion for the quantum state is the Schrodinger equation, a partial differential equation defined as
\begin{equation}
\label{eq:schro}
i\hbar \frac{\partial \ket{\Psi(t)}}{\partial t} = \widehat{H}(t) \ket{\Psi(t)}
\end{equation}
where the time-dependent operator $\widehat{H}(t)$ defines the energetic interactions governing the physical system, and is referred to as the \emph{Hamiltonian}. Consequently, the Hamiltonian is important for manipulating the quantum state and its control plays a prominent role in the design and testing of quantum computing technologies. It is important to note that a quantum state can not be directly observed by physical measurement. Rather measurements of a quantum state must be performed relative to a basis set, e.g., $\{\ket{j}\}$. The probability to observe the $i$-th outcome corresponds to the probability $p_i$ defined above, such that a series of repeated measurements over an ensemble of identically prepared quantum states will generate a distribution of outcomes that approximates the set of probabilities $\{p_j\}$. Thus, the accurate characterization of this distribution can be exceedingly difficult due to the large number of basis states and the infrequent occurrence of measurement outcomes corresponding to low probabilities. A survey of methods for measuring quantum state is provided in Ref.~\cite{Holevo2011}
\begin{figure}[ht]
\centering
\includegraphics[width=0.5\columnwidth]{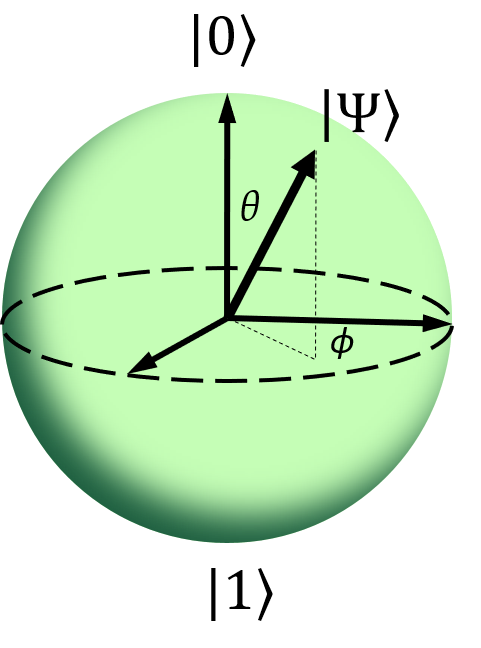}
\caption{The Bloch sphere with a unit radius provides a geometrical representation of a qubit. The north and south poles of the sphere define the orthonormal basis states $\ket{0}$ and $\ket{1}$, respectively, while the surface defines the set of all possible qubit values. In spherical coordinates, the example qubit $\ket{\Psi}$ has expansion coefficients $c_0 = \cos\theta$ and $c_1 = e^{i \phi} \sin{\theta}$.}
\label{fig:qubit}
\end{figure}
\par
A prominent example of a quantum state within the context of quantum computing is the case of a \emph{qubit}. A qubit, or quantum bit, refers to the quantum state of an isolated two-level quantum mechanical system. Informally, the qubit is the quantum analog of bit. A qubit is the fundamental unit of information within quantum computing. In the development of quantum computing technologies, a qubit is stored within a physical two-level system. We denote those physical systems as quantum register elements, in which an individual quantum register element represents the ability to store a single qubit of information. We will discuss some of the different physical systems as quantum register elements in Sec.~\ref{sec:devices}. Logically, the qubit is defined over a basis of binary states labeled as `0' and `1', respectively, such that an arbitrary state of a qubit may be expressed as the linear combination
\begin{equation}
\ket{\psi} = c_0 \ket{0} + c_1 \ket{1}
\end{equation}
The \emph{superposition} of orthogonal basis states is fundamental to quantum mechanics. Recall that the expansion coefficients are complex-valued numbers normalized as $|c_0|^2 + |c_1|^2 = 1$. As the absolute phase of a quantum state is arbitrary \cite{Sakurai1995}, a convenient graphical representation of the qubit is given in spherical coordinates. As shown in Fig.~\ref{fig:qubit}, the surface of a unit sphere represents all possible qubit values, where the points of $\ket{0}$ and $\ket{1}$ are located at the north and south poles, respectively. An arbitrary quantum state $\ket{\Psi}$ is normalized to unity and must lie on the surface of the sphere. In Fig.~\ref{fig:qubit}, the amplitudes $c_0$ and $c_1$ represent the projection of the quantum state onto the corresponding basis states and the example qubit $\ket{\Psi}$ has expansion coefficients $c_0 = \cos\theta$ and $c_1 = e^{i \phi} \sin{\theta}$. This representation of the qubit state on a unit sphere is commonly called the Bloch sphere in quantum mechanics.  
\par
A multi-qubit register is an addressable array of $n$ two-level physical systems. The principle of superposition may be extended to the register as the quantum state for the composite physical system is also given by Eq.~(\ref{eq:psi}). For an $n$-qubit register, the computational basis is expressed in binary notation as
\begin{equation}
\ket{j} = \ket{j_1,j_2,\ldots,j_n} = \ket{j_1}\otimes\ket{j_2}\ldots\otimes\ket{j_n},
\end{equation}
where the binary values $j_{k}$ correspond to the binary expansion of $j$. The dimensionality of the underlying Hilbert space is $N = 2^{n}$ and any normalized vector represents a valid quantum state. In particular, there are composite quantum states which cannot be expressed as separable products of $n$ single-qubit states. Such states are known as \emph{entangled} states and they are a hallmark of quantum mechanics and, therefore, quantum computing. For example, consider the quantum state of a 2-qubit register
\begin{equation}
\ket{\Psi} = \frac{1}{\sqrt{2}}\left(\ket{0,0} + \ket{1,1} \right).
\end{equation}
Measuring the individual elements of the register will generate binary outcomes 0 or 1 with equal probability. Accordingly, the classical expectation for a joint measurement of the register is a uniform distribution of four possible outcomes. However, measurements of this quantum state are always correlated such that both results are either (0,0) or (1,1), where the probability for each of these outcomes is $1/2$. Notably, there is no possibility for observing anti-correlated outcomes for this quantum state, e.g., (0, 1). The presence of these correlations in the measurement statistics is known as \emph{entanglement} and the underlying quantum state is said to be entangled. Fundamentally, entanglement is a limitation on the ability to describe states of a register solely by specifying the value of each register element, and entangled states are notable for the ability to violate the local, causal relations predicted by classical mechanics \cite{Horodecki2009}.
\par
The \emph{no-cloning principle} represents a fundamental constraint placed on quantum information processing. The no-cloning principle is a consequence of the linearity of quantum mechanics \cite{Wootters1982}, in which the ability to perfectly clone, aka copy, an arbitrary quantum state is not permitted. In particular, given a quantum register storing an arbitrary state $\ket{\Psi_1}$, this information cannot be copied into a second register without loss of information. Efforts to optimally approximate the value of the first register, known as quantum cloning \cite{Scarani2005}, can be evaluated by measuring the fidelity defined 
\begin{equation}
f = |\inp{\Psi_2}{\Psi_1}|^2,
\end{equation}
where $\ket{\Psi_2}$ is the value of the second register and $f \in [0,1]$.
\par
The principles of operation for a quantum computer are based on the Schrodinger's equation in Eq.~(\ref{eq:schro}), in which the time-dependent Hamiltonian $\widehat{H}(t)$ can be directly controlled through the use of externally applied fields. Depending on the specific technology in place, these controls will consist of electrical, magnetic, or optical fields designed to drive the dynamics toward a specific response. In Sec.~\ref{sec:devices}, we present examples for devices based on semiconductors, superconductors, and trapped ion technologies. In some computational models, the time-dependent controls are realized as pulsed fields that act discretely on the quantum register elements. These discrete periods of field interaction are known as \emph{gates} and the effect of the gate on the quantum register is described by an unitary operator that transforms the stored quantum state. This is known as the \emph{gate} or \emph{circuit model} since a diagrammatic sequence of gates acting on registers provides a design for instruction execution. 
\par
An alternative computational model applies the time-dependent field as continuous interaction subject to constraints on the rate of change for the overall Hamiltonian. This constraint imposes an adiabatic condition on the dynamics of the quantum system \cite{Amin2009}, such that the Hamiltonian slowly modifies the interactions between quantum physical subsystems, i.e., register elements, relative to the internal energy scales describing those subsystems. As a result, the register state can be driven toward a desired outcome. This is known as the \emph{adiabatic model} given the constraints on the controls. A device design based on the adiabatic model has been implemented in superconducting technology by the commercial vendor D-Wave Systems, Inc. In the realization of that design, the Hamiltonian control is restricted to a specific functional form, namely the transverse Ising model, which limits the device operation to computing discrete optimization problems. In addition, the physics of the device are not well modeled by the Schrodinger equation, cf.~Eq.~(\ref{eq:schro}), but rather require a more sophisticated model that includes non-trivial interactions with the surrounding quantum physical systems as well as finite temperature effects \cite{Johnson2011}. Nonetheless, the device has been observed to correctly compute the solution to a wide variety of discrete optimization problems and has been characterized as having some advantages relative to conventional computing devices. While the remainder of this tutorial will focus on the gate model for quantum computing, we refer the reader interested in adiabatic quantum computing to the recent review by Albash and Lidar \cite{Albash2018}.
\par
We now summarize the \emph{basic criteria} that define the expected functionality of quantum computing devices. As first presented by DiVincenzo \cite{DiVincenzo2000}, these criteria represent the minimal behaviors needed to perform general-purpose quantum computing in the presence of likely architectural constraints. First is the ability to address the elements in a scalable register of quantum systems. Scalability implies a manufacturing capability to fabricate and layout as many register elements as needed for a specific computation. Second, these register elements must be capable of being initialized with high fidelity, as the starting quantum state of the computation must be well-known to ensure accurate results. The third criterion is the ability to measure register elements in a well-specified basis. As discussed above, measurement samples the statistical distribution encoded by the quantum state according the probabilities $p_{i}$ over a given basis set. A measurement sample represents readout from the register of the quantum computer and this value may be subsequently processed. 
\par
Fourth, the control over the register must include the ability to apply sequences of gates drawn from a universal set. Universality of the gate set characterizes the potential to perform an arbitrary unitary operation on the quantum state using a sufficiently long series of gates from that set. In particular, it is known that a finite set of gates is sufficient to approximate universality and, moreover, that a finite set of addressable one- and two-qubit gates are sufficient for universality \cite{Dawson2005}. The latter result, known as the Solovay-Kitaev theorem, provides a constructive method for composing arbitrary gates from a finite, universal gate set. Selection of a universal gate set raises the question of the optimal instruction set architecture for an intended application within a specific device technology \cite{Britt2017instruction}. The fifth criterion is that the gate operation times must be much shorter than the characteristic interaction times on which the register couples to other unintended quantum physical systems. These interactions induce decoherence of the stored quantum superposition states, which leads to the loss of information \cite{Schlosshauer2005,Streltsov2017}. In order to maintain the stored quantum state with sufficient accuracy, the duration of the gate sequence must be shorter than the characteristic decoherence time. Fault-tolerant protocols for gate operations are designed to counter the losses from decoherence and other errors by redundantly encoding information with quantum error correction codes \cite{Campbell2017}.
\par
Two additional functional criteria are necessary for a quantum computer with geometrical constraints on the \emph{layout} of the quantum register. In particular, layout constraints may impose restrictions on which register elements can be addressed by multi-qubit gates, e.g., nearest neighbors within a two-dimensional rectangular lattice design. Physical layout restrictions may be overcome by moving stored quantum states between register elements. This is accomplished using the SWAP gate, a unitary operation that exchanges the quantum state between two register elements. In addition, a MOVE operation can support long distance transport of a stored value, in which the register element itself is displaced. The latter proves useful for distributed quantum registers that may requires interconnects, aka communication buses, to SWAP register values. The necessity of these functions depends on the purpose of the quantum computer and especially the limitations of the technology. Presently, all technologies for quantum computing face some constraints on register layout.

\section{Devices for Quantum Computing}
\label{sec:devices}
%
There are many different possible technologies available for building quantum computers, and these are typically classified by how qubits of information are stored. As discussed in Sec.~\ref{sec:principles}, these devices must meet several functional criteria to carry out reliable quantum computation. In this section, we provide an overview of three technologies that are currently used for developing quantum computing devices and we discuss the progress toward meeting the functional criteria. 
\subsection{Silicon Spin Qubits}
\begin{figure*}[t]
\centering
\includegraphics[width=\textwidth]{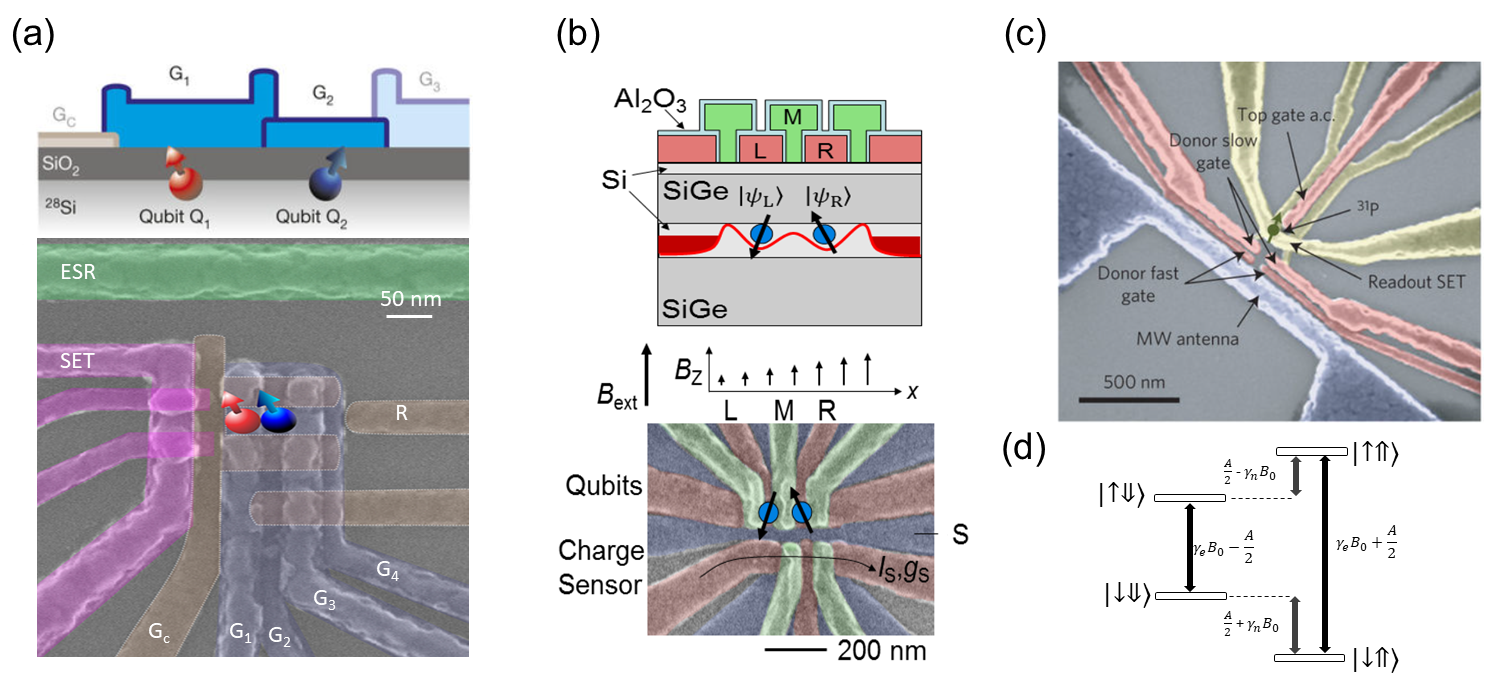}
\caption{(a) Bottom panel: Scanning electron microscope (SEM) image of a metal-oxide-semiconductor (MOS) quantum dot device similar to the one where single and two-qubit gates were demonstrated. Top panel: Cross-sectional schematic of the device illustrating the location of qubits at the Si/SiO2 interface. \href{https://www.nature.com/articles/nature15263}{This Figure is reprinted from Ref. \cite{Veldhorst2015N} with permission from Nature}. (b) Bottom panel: SEM image of a Si/SiGe double quantum dot device, where two-qubit operations were implemented. Middle panel : Variation of the static magnetic field along a slice of the device. Top panel : Cross sectional device schematic highlighting
the position of the quantum dots. \href{http://science.sciencemag.org/content/359/6374/439}{This Figure is reprinted from Ref. \cite{zajac2018science} with permission from the American Association for the Advancement of Science (AAAS)}. (c) SEM image of an ion-implanted $^{31}$P device similar to the one used for demonstrating  record spin-coherence times \cite{laucht2017N, Muhonen2014nn}. (d) $^{31}$P donor electron ($\lvert \uparrow\rangle$, $\lvert \downarrow\rangle$) and nuclear ($\lvert \Uparrow\rangle$, $\lvert \Downarrow\rangle$) spins states with the relevant energy separation between them \cite{humble2016nanotech}.}
\label{fig:SiliconQubitPlatforms}
\end{figure*}

Silicon spin qubits denote a technology implementation by which quantum information is encoded either in the \emph{spin states} of an electron found in a silicon quantum dot, or in the spin state of the electron or nucleus of a single-dopant atom (typically group V donors) in a silicon substrate. In particular, the orientation of the spin in these systems is used to encode the $\ket{0}$ and $\ket{1}$ states. Notably, these silicon devices are fabricated with conventional CMOS techniques, and consist of gate electrodes (normally Aluminum or Polysilicon) that can control the energy landscape in the silicon substrate. These electrodes are appropriately designed and biased such that a single electron is confined in a quantum dot at the interface. Examples of a silicon quantum dot include the MOS device shown in Fig.~\ref{fig:SiliconQubitPlatforms}(a) or the Si/SiGe device shown in Fig.~\ref{fig:SiliconQubitPlatforms}(b). Similar electrostatic control is used for silicon donor devices like the example shown in Fig.~\ref{fig:SiliconQubitPlatforms}(c) of a Phosphorus donor implanted inside a silicon substrate. In all of these examples, the electrons are strongly confined such that lowest electronic orbital energy in the quantum dot or the donor is well isolated from other excited electronic states. The confinement length for the donor electron is $\sim1.5$ nm in all 3-dimensions, while for the dot electron, these dimensions are $\sim10$ nm  and $\sim2$ nm in the lateral and vertical directions, respectively. These characteristic dimensions make silicon qubits the most compactly fabricated technology as compared to the qubit technologies discussed in later subsections. 
\par
Addressing silicon spin qubits uses an applied static magnetic field $B_0$ to split the orbital degeneracy of the dot-electron at the interface. Due to the \emph{Zeeman effect}, the orbital for the confined electron is split into the distinct spin states  $\lvert \uparrow\rangle$ and $\lvert \downarrow\rangle$. These spin states encode the computational states $\ket{0}$ and $\ket{1}$, where the energy splitting is given by the Zeeman energy $\gamma_eB_0$ with $\gamma_e \sim 28$ GHz/T the gyromagnetic ratio of the electron. For $^{31}$P donors, the  electron and nuclear spins are coupled by the \emph{hyperfine interaction} $A \sim 117$ MHz \cite{Feher1959pr}. The donor qubits are generally operated under large magnetic fields $B_0 > 1$ T, such that $\left(\gamma_e + \gamma_n\right)B_0 \gg A$, where $\gamma_n \sim 17$ MHz/T is the gyromagnetic ratio of the nucleus.  In this limit, the eigen spin states are tensor products of the electronic spin ($\lvert \uparrow\rangle$, $\lvert \downarrow\rangle$) and the nuclear spin ($\lvert \Uparrow\rangle$, $\lvert\Downarrow\rangle$) states. The resulting energies are shown in Fig.~\ref{fig:SiliconQubitPlatforms}(d), where the electron spin qubit splitting depends on the nuclear spin states, and vice versa. Typical energy splittings are of the order of tens of GHz and MHz for the electron and nuclear spins, respectively \cite{Pla2012n,Pla2013n}. The hyperfine interaction $A$ and the electron gyromagnetic ratio $\gamma_e$ depend on the orbital wavefunction of the electron, which can be tuned with electric fields \cite{Laucht2015sciadv, veldhorst2015spin}. As a result, the qubit splittings are electrically tunable after the silicon qubit devices are fabricated.
\par
Electron spin qubits are commonly initialized and measured using \emph{spin-charge conversion} techniques \cite{Morello2010n}. Charge sensors such as quantum point contacts and single-electron-transistors (SET) are located adjacent to the quantum dot (or donor) and are then capacitively coupled to them, cf. Fig.~ \ref{fig:SiliconQubitPlatforms}. The charge sensors are biased appropriately with gate voltages, such that the current passing through them is strongly sensitive to the electrostatic environment in their vicinity. The orbital energy of the electron is then electrically tuned such that the electron can preferentially tunnel to the same or another nearby charge reservoir, depending on its spin. The presence or absence of the electron on the donor/dot can then be detected via a change in current passing through the charge sensors, which aids to readout the electron spin state. The protocol will also initialize the electron spin state in the dot or the donor to $\lvert \downarrow \rangle$ \cite{Morello2010n}.
\par
For spin control, an oscillating (driving) magnetic field is applied to the qubits. The frequency of the oscillating field is chosen to be equivalent to the energy difference between the two spin qubit levels. Based on the principles of \emph{magnetic resonance}, transitions between the spin states are then achieved at a rate proportional to the amplitude of the driving field \cite{slichter2013principles}. The driving field is pulsed appropriately to obtain a specific rotation of the spin state, for implementing a single qubit gate like the Hadamard gate. A microwave transmission line antenna (see Fig.  \ref{fig:SiliconQubitPlatforms}a and Fig. \ref{fig:SiliconQubitPlatforms}c) is normally used to generate the driving field \cite{Dehollain2013nt}, yielding magnetic field amplitudes of $\sim 0.1$ mT, and single qubit gate times of few micro-seconds \cite{Pla2012n} (or milliseconds \cite{Pla2013n}) for the electron (or nucleus). Alternatively, a micromagnet producing a dc magnetic field gradient (Fig.  \ref{fig:SiliconQubitPlatforms}b) can also be embedded on chip \cite{kawakami2016pnas}. In the presence of an additional oscillating electric field (from gate voltages), the electron feels an effective oscillating magnetic field, resulting in spin-resonance with faster gate times. Note that the frequency of the control field is different for both the electron (ESR frequencies $\sim$ tens of GHz) and the nucleus (NMR frequencies $\sim$ tens of MHz). The ability to control and readout the electron spin state also allows measurement of the nuclear spin state. As the electron spin resonance frequency is determined by the nuclear spin state (see Fig.  \ref{fig:SiliconQubitPlatforms}d), probing frequencies at which the electron can be controlled, allows readout of the nuclear spin \cite{Pla2013n}.
\par
Since the splittings are dependent on $A$ and $\gamma_e$, they can be tuned electrically and it is possible to independently control each donor located within a precisely positioned array \cite{Kane1998n}. In their idle state, the qubits are electrically detuned from the control field by appropriately tuning $A$ and $\gamma_e$. When operations need to be performed on the qubits, they are brought in resonance with the control field, i.e. the energy splitting is tuned to the frequency of the control field. 
\par 
The coupling between two electron spin qubits occurs via the intrinsic \emph{exchange interaction} between them \cite{Kane1998n}. The exchange coupling $J_e$ is primarily determined by the overlap between the two-electron wave functions.  $J_e$ can hence be tuned by either modifying the tunnel barrier between the two electrons, or by shifting the relative orbital energies of the two electrons \cite {shim2017arxiv}. Both these methods can be realized by appropriately tuning the gate voltages that control the potential landscape in the device. To perform a CNOT gate, the electron spin qubits are operated in a regime where $J_e$ is smaller than the energy difference between the qubit splittings of the two electrons (often termed as the detuning). In such a regime, each electron spin qubit will have two resonance frequencies, which are determined by the state of the other qubit. Hence, an oscillating control field at one resonant frequency will conditionally rotate the qubit dependent on the state of the other qubit, resulting in a CNOT gate \cite{Veldhorst2015N,zajac2018science}. To perform SWAP, the qubits are initialized in a regime, where the exchange coupling is much smaller than their detuning. The exchange coupling is then increased to a value much larger than their detuning, such that the two qubits exchange information with each other. After an appropriate time that determines the angle of SWAP, the exchange coupling is brought back to a low value. 
\par
The \emph{spin-orbit coupling} is weak for  electrons in silicon, resulting in long spin-relaxation times $T_1$. The relaxation time has been shown to be dependent on the temperature and magnetic field \cite{zwanenburg2013rmp}. Operating the qubits at low temperatures ($<$ 1 K) and magnetic fields ($<$ 5 T), yield $T_1$ exceeding several seconds and even hours. The presence of spin containing nuclei (such as Si-29) in the lattice, and their fluctuations, can result in decoherence of the electron spins \cite{Witzel2010prl}. Hence, \emph{isotopic purification} of silicon from spin containing nuclei, allows for long-coherence times ($T_2$) of milli-seconds and seconds for the electron and nuclear spins respectively  \cite{Muhonen2014nn}. Additional sources of decoherence include charge or electric field noise arising from nearby defects/traps, control signals, gate electrodes and thermal radiation from the microwave antenna \cite{Muhonen2014nn}.
\par
While the methods to address and couple silicon qubits can be integrated with the microelectronics industry, the qubits are very sensitive to \emph{atomic details} that have not yet been addressed in the industry. These details strongly affect the qubit operation, and hence it is essential to design devices that minimizes their influence on the qubits. First, the exchange coupling between  donor electrons is extremely sensitive to the position of donors, necessitating \emph{precise donor placement} accuracies and/or large exchange coupling tunabilities \cite{Koiller2002prl,song2016apl}. Efforts are underway to demonstrate qubits with  single-donor atoms in silicon that are placed precisely with Scanning Tunneling Microscopy \cite{Fuechsle2012nn}, as well as to explore alternate means of coupling between the qubits (such as dipolar interactions \cite{tosi2017nc, hill2015sciadv}) that are less sensitive to donor placement inaccuracies. In addition, \emph{atomic roughness} and \emph{step edges} at the interface, can result in the excited orbital states coming close to the ground orbital state in silicon quantum dots, accelerating relaxation and even resulting in a non spin-1/2 ground states \cite{zwanenburg2013rmp}. The energy separation between the ground and excited orbital states (also referred to as valley splitting) can be tuned with electric field to an extent \cite{yang13nc}, yet it is always desirable to obtain larger and uniform valley splittings with a smooth interface. Finally, uncontrolled \emph{strain} in the lattice arises from the thermal mismatch between the gate and substrate materials when the device is cooled from room temperature to milli-Kelvin temperatures \cite{thorbeck2015formation}. This modifies the potential landscape in the device, altering the position and confinement of the quantum dots, along with introducing accidental dots. Ref.~\cite{thorbeck2015formation} highlights that using gate materials (such as polysilicon rather than aluminum) which have similar thermal expansion coefficients to that of silicon, can aid to reduce the lattice strain. 
\par
The exchange interaction between the qubits is short-range (within few tens of nm), and can only result in nearest-neighbor couplings. To scale up silicon qubit devices to a large-scale architecture, it is beneficial to have connectivity between qubits that are separated by much larger distances. Methods to couple silicon qubits to a \emph{photonic mode} spanning $\sim$ centimeter in a microwave resonator have been proposed previously \cite{tosi2017nc,hu2012prb}, and recently demonstrated in Si/SiGe quantum dots \cite{samkharadze2018science, mi2016arxiv}. Through the photonic mode, two qubits separated by as far a centimeter can be virtually coupled to each other, enhancing the qubit-connectivity significantly. Coupling the spins to the resonator also provides a pathway to readout the spin states \cite{tosi2017nc}. The transmission frequency of the resonator then depends on the spin state of the qubit. Hence, applying a microwave signal to the resonator, and measuring its transmission aids to detect the spin state.
\par
Designing silicon spin qubit devices requires modeling several classical and quantum mechanical parameters with a range of techniques that are adapted from the semiconductor industry \cite{humble2016nanotech}. Classical variables that are relevant and need to be solved for include the electrostatic potential landscape, electric fields, electron densities, capacitances, magnetic fields and strain. The electrostatic parameters in silicon devices can be obtained by solving Poisson's equation with the finite-element method with traditional TCAD design packages such as Sentaurus TCAD, or a general multiphysics package like COMSOL. Solving Maxwell's equations with high-frequency electromagnetic solvers (such as CST-Microwave Studio or ANSYS-HFSS) aids to estimate the driving magnetic fields generated by the microwave antenna in such devices. Thermal strain while cooling such devices can also be simulated by solving the stress-strain equations with COMSOL \cite{ thorbeck2015formation}. In addition to the classical parameters, it is also essential to solve the electronic structure in silicon qubit devices, and estimate the electron orbital-energies, and wave functions. Effective mass theory and tight-binding techniques have been extensively used for such calculations \cite{zwanenburg2013rmp}. The orbital energies and wave functions act as a handle to the hyperfine, exchange  and tunnel couplings, along with the electron gyromagnetic ratio and electron spin relaxation times. These parameters are ultimately fed into a simplified spin Hamiltonian, which is solved with mathematical packages (such as MATLAB, Mathematica or QuTiP), to simulate the instantaneous spin states and quantum gate fidelities.
\subsection{Trapped Ion Qubits}
\begin{figure*}[ht]
\centering
\includegraphics[width=\textwidth]{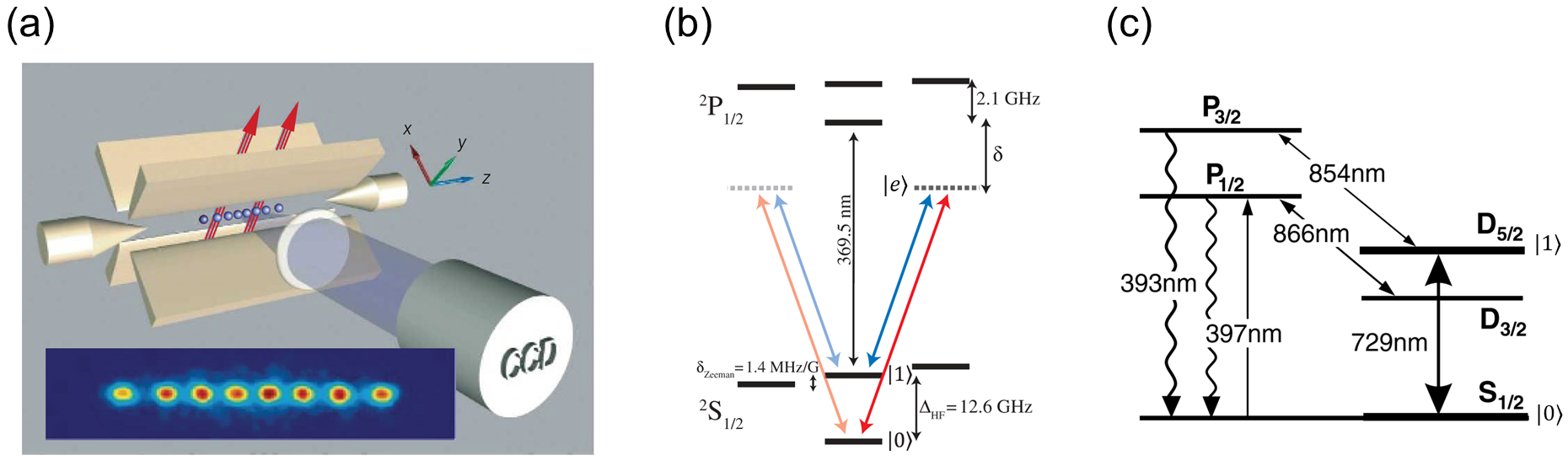}
\caption{(a) Schematic of a Paul trap used to confine ions in vacuum. Inset : Visualization of ions in the trap with fluorescent techniques. \href{https://www.nature.com/articles/nature07125}{This Figure is reprinted from Ref. \cite{blatt2008n} with permission from Nature}. (b) Electronic energy levels of a $^{171}$Yb$^+$ ion illustrating qubit encoding ($\lvert0\rangle$ and $\lvert1\rangle$) with hyperfine energy levels \cite{mount2013njp}. Transition between qubit  states is achieved by a Raman process via excitation to a virtual state $\lvert e\rangle$. (c) Electronic energy levels of a $^{40}$Ca$^+$ ion illustrating qubit encoding with the s- and d-orbital energy levels. \href{https://link.springer.com/article/10.1007\%2Fs11128-004-3105-1}{This Figure is reprinted from Ref. \cite{blatt2004qip} with permission from Springer}.}
\label{fig:IonTrapQubitPlatforms}
\end{figure*}

Trapped ion qubits represent an implementation where quantum information is encoded in the \emph{electronic energy levels} of ions suspended in vacuum. To obtain trapped ions, metals such as Calcium (Ca) or Ytterbium (Yb) are first  resistively heated and vaporized with a current passing through them, and then directed to the trap. While loading these ions into the trap, these vaporized neutral atoms are simultaneously photo-ionized, where their outermost electron is removed, resulting in ions that have a single valence electron. As the ions are charged particles, appropriate voltages applied to gate electrodes in their vicinity and resulting electric fields, can then confine the ions in the trap. The most common gate-electrode configuration for ion trapping is the \emph{(rf) Paul trap} (Fig.  \ref{fig:IonTrapQubitPlatforms}a), which consists of 4 electrodes (2 with oscillating voltages and 2 grounded) that induce an effective harmonic potential in the x-y plane, and additional two DC gate electrodes to induce harmonic confinement in the z-plane \cite{paul1990rmp}. In the harmonic oscillator potential, there are several eigen states corresponding to the vibrational modes of the trapped ions. To ensure that thermal effects and fluctuating electromagnetic fields do not cause random excitation of these states and thereby motion of the ions, the ions are laser-cooled to their vibrational ground state \cite{leibfried2003quantum}. For a small number of ions ($\sim 50$), the ions will then be arranged in a linear chain along the z-direction, such that overall forces from the external fields cancel out the forces from their Coulomb interaction. Typical ion separation in the trap is $\sim 10 $ $\mu$m.
\par
As mentioned previously, a qubit is defined using the energy levels of individual ions in the trap to encode the basis states $\lvert 0\rangle$ and $\lvert1 \rangle$. Depending on the orbital energy levels used for encoding, there are two
popular implementations of trapped ion qubits : \emph{hyperfine} and \emph{optical} qubits. For hyperfine qubits, the states correspond to the hyperfine levels in the atomic s-orbital. For example, as shown in Fig.~ \ref{fig:IonTrapQubitPlatforms}(b), the ion $^{171}$Yb$^+$ has a nuclear spin of 1/2, and the qubit is encoded using the singlet $\lvert S \rangle$ and $\lvert T_0 \rangle$ configurations of the electron and nuclear spins \cite{blinov2004qip}. A small DC magnetic field is applied to separate the $\lvert T_0 \rangle$ state from other triplet states $\lvert T_- \rangle$ and $\lvert T_+ \rangle$. The qubit splitting of 12.6 GHz for $^{171}$Yb$^+$ is determined by the hyperfine interaction between the electron and the nucleus, and insensitive to magnetic field fluctuations up to first order \cite{fisk1997ieee}. Alternatively, for the optical qubit encoding with trapped ions, the basis corresponds to s-orbital and d-orbital electronic energy levels. As shown for $^{40}$Ca$^+$ in Fig. ~\ref{fig:IonTrapQubitPlatforms}(c) \cite{blatt2004qip}, the energy splitting is then $\approx 411$ THz and equivalent to 729 nm. Trapped ion qubits are highly reproducible \cite{ludlow2015rmp} provided there are no magnetic and electric field inhomogeneities in the trap, which may modify the energy levels through Stark and Zeeman effects respectively.
\par
Fluorescent techniques are used to visualize the ions, where the qubit states are continuously excited to the p-states with the help of a laser, to induce an \emph{electric-dipole transition} \cite{leibfried2003quantum}. On such a transition, the ions scatters the photons which are detected by photo-multiplers or a CCD camera (see Fig.  \ref{fig:IonTrapQubitPlatforms}a). The required laser frequency is equivalent to the separation between the energy states used for the transition, and depends on the choice of the ion.
\par
The hyperfine and optical qubits are initialized with \emph{optical pumping}. Here, a laser is incident on the ions with an appropriate frequency that can continuously drive the $\lvert 1 \rangle$ state to the excited p-states. Any spontaneous decay from the excited p-state to ground states apart from $\lvert 0 \rangle$, are also further driven by the laser \cite{debnath2016thesis}. Over a period of time ($\sim \mu$s), all the spontaneous emissions result in the qubit state being initialized to $\lvert 0 \rangle$ \cite{monroe2013s}.
\par
For readout of trapped ion qubits, the laser is tuned to a frequency that continuously drives one of the basis states (e.g. $\lvert 1 \rangle$) to an excited p-state. The polarization of the laser and excited state is chosen such that spontaneous emission cannot occur to the other basis state $\lvert 0 \rangle$, based on spin-selection rules \cite{debnath2016thesis}. Hence, if the initial qubit state is $\lvert 1 \rangle$, the resulting p-state after excitation may spontaneously decay to states apart from $\lvert 0 \rangle$, which are also continuously excited. Photons from the spontaneous emission are then detected with a CCD camera. If the initial qubit state is $\lvert 0 \rangle$, the qubit cannot be excited to the p-states by the laser, as its frequency is far away from resonance, and there is no output at the CCD-camera.
\par
For optical qubits, a stable laser (having $\sim$ 400 THz frequencies) with a narrow line-width can drive the transitions between the $\lvert 0 \rangle$ and $\lvert 1 \rangle$ states via a \emph{quadrupole transition}, enabling qubit control \cite{haffner2008pr}. The hyperfine qubits can be controlled with two methods. (i) Microwave radiation with frequencies (e.g. $12.6$ GHz for $^{171}$Yb$^+$) matching the qubit splitting can drive transitions between $\lvert 0 \rangle$ and $\lvert 1 \rangle$ states \cite{blume2017nc}. Microwaves can be generated with a microwave horn that is located several centimeters from the trap. However, since microwaves correspond to centimeters in wave length, and the ions are separated by micrometers, it is not possible to focus microwaves and address individual qubits in a chain of several ions. (ii) Alternatively, \emph{stimulated Raman transitions} with  two laser fields (from pulsed laser) can be used to control the qubit state \cite{brown2016npj}. Each laser field excites the qubit states to a virtual level $\lvert e \rangle$ that is well detuned (by $\delta$) from the excited p-states (see Fig.  \ref{fig:IonTrapQubitPlatforms}b). The frequency difference between the two laser fields is chosen to match the qubit splitting. Based on a Raman process, the spins are rotated at a frequency proportional to the product of the individual Rabi frequencies (from $\lvert 0 \rangle$ to  $\lvert e \rangle$⟩ and from $\lvert 1 \rangle$ to $\lvert e \rangle$⟩ determined by the laser power), and inversely proportional to the detuning $\delta$ from the p states. This method has the advantage of selectively addressing the qubits, where the laser can be focused individually on each qubit.  Typical timescales for single qubit operations are of the order of several microseconds.
\par
The Coulomb interaction between the ions serves to mediate the coupling between the qubits \cite{blatt2008n}. Based on this interaction, the qubit states are coupled to the \emph{vibrational modes} of the ion-chain. Hence, appropriate laser frequencies can help to transfer the qubit states to the vibrational modes. Depending on the vibrational modes of the ion-trap, a subsequent ion in the chain can be rotated  with a laser, to demonstrate a CNOT gate. The vibrational modes can also be swapped with the subsequent qubit, resulting in a SWAP gate.

\par
Like silicon spin qubits, trapped ion qubits have extremely long relaxation and coherence times. The relaxation mechanism is via spontaneous decay which approach several seconds for optical qubits, and several days for hyperfine qubits. The coherence of the qubits is primarily affected by ambient magnetic field fluctuations which modify the qubit energy levels through the Zeeman effect, laser intensity and frequency fluctuations over time, and coupling of the qubit states to the vibrational degree of freedom during 2-qubit operations \cite{wineland1998nist}. The sources of decoherence for the vibrational degree of freedom include unstable trap parameters, coupling of the electric dipole associated with the motion of ions to thermal radiation in the environment, and ion collisions with the residual background gas. Typical coherence times of the trapped ion qubits due to these effects is of the order of seconds.   

\par
The coupling rate between the qubit state and vibrational mode (for two qubit operations) has been shown to be inversely proportional to the square root of the number of ions in the chain \cite{monroe2013s}. Hence, increasing the ion number in the chain beyond $\sim$ 50 slows down the 2-qubit operations, where decoherence (heating) of the motional modes and fluctuating electric fields become significant. Architectures for scale up with larger number of ions include \emph{Quantum Charge Coupled Device} (QCCD) architectures \cite{kielpinski2002nature} where individual ions at the edges of a trap are shuttled to nearby traps and made to interact with them, for connecting distant qubits. This would require exquisite control of the shuttling of the atomic ions, as well as the periodically cooling down the excess motion arising from shuttling ions. While this method could potentially work for larger number of qubits ($\sim$ 1000), it becomes impractical for scale-up  due to complexity of interconnects, diffraction of optical beams, and extensive hardware requirements. \emph{Photonic interfaces} have been proposed to connect even larger systems \cite{monroe2013s}. Here, qubits at the edges of the chain are driven to an excited state with very fast laser pulses so that at most one photon emerges from each qubit after radiative decay. Following selection rules, the radiative decay can lead to entanglement between the photonic and trapped ion qubit. Photons from two separate qubits are mode-matched and interfered on a beam-splitter, which is then detected. A successful detection then yields an entangled state between the two distant ion trap qubits.

\par
The design packages available in the conventional microelectronics industry cannot be directly extended to design trapped ion qubits, as their implementation has very little overlap with that of silicon. Nevertheless, the electric fields available from classical electrostatic solvers (such as COMSOL) can be used to optimize and design the gate electrode configuration and voltages for the trap. As illustrated previously in this section, the electronic orbital levels of single ions (or even a cluster of ions) in the trap, determine the laser frequencies needed for initialization, readout, control and coupling of the trapped ion qubits. The orbital energies and hyperfine interactions for a variety of trapped ion candidate materials can be determined from ab-initio electronic structure calculation techniques such as density functional theory (DFT). A significant aspect of the design also include the optical setup for the lasers, including its power and focus. These parameters can be obtained with commercial ray-tracing software packages such as Zemax, Code V or Oslo. The dynamics of the trapped ion qubits upon interaction with a laser can be mapped onto a simplified Hamiltonian, which can then be solved with commercial mathematical packages, such as MATLAB. While there are several analytical expressions and mathematical models for light-matter interactions, a device simulator capable of capturing the non-idealities in realistic trapped ion devices is currently non-existent.           

\subsection{Superconducting Transmon Qubits}

\begin{figure*}[t]
\centering
\includegraphics[width=\textwidth]{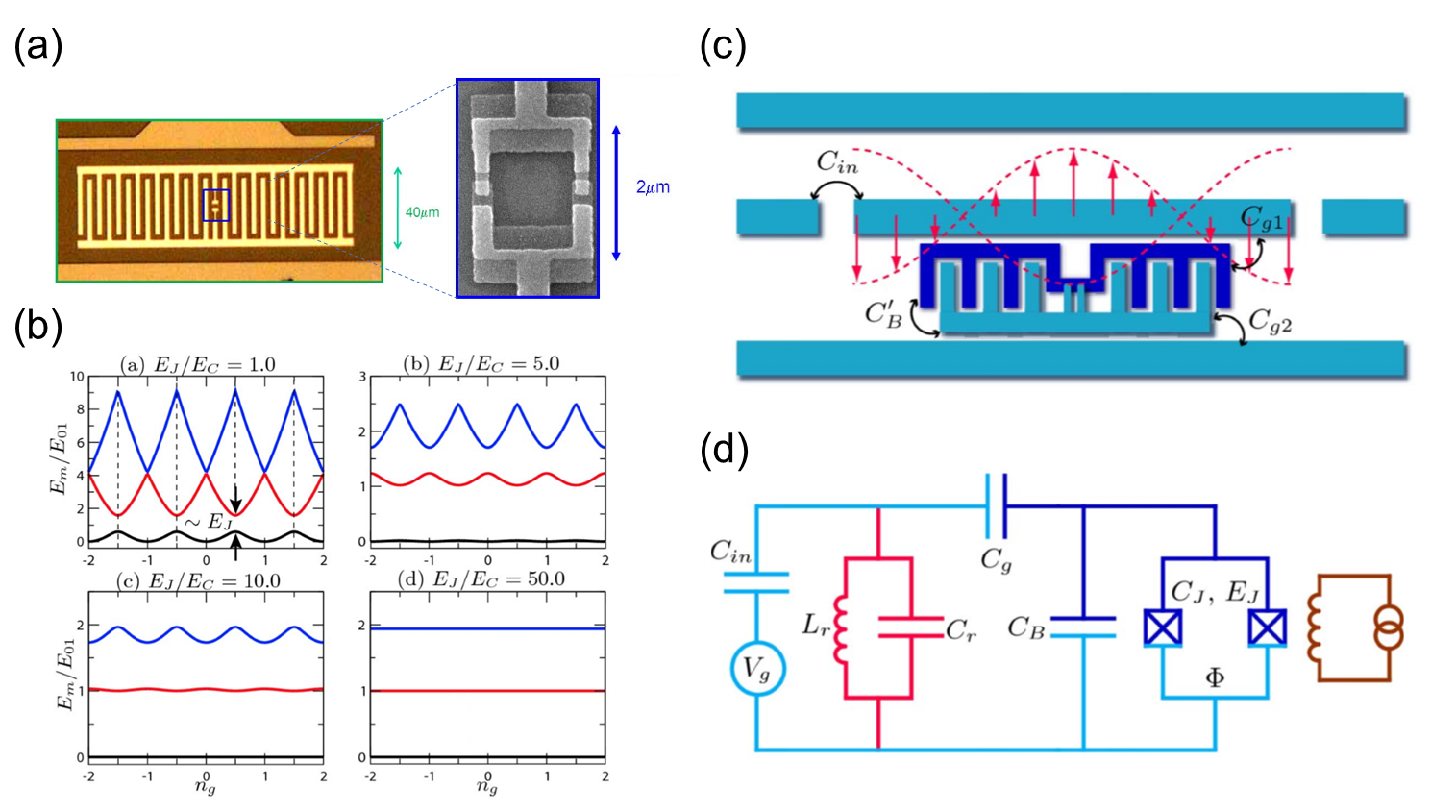}
\caption{(a) The transmon qubit consisting of two superconducting islands that are coupled through Josephson junctions and a large interdigitated capacitance. Inset : SEM image of the device in the vicinity of the Josephson junctions. (b) Eigenenergies $E_m$ (first three levels, $m = $ 0, 1, 2) of the superconducting system as a function of the effective offset charge $n_g$ induced by nearby gate electrodes and environment  \cite{koch2007pra}. Energies are given in units of the transition energy $E_{01} = E_1 - E_0$ evaluated at $n_g = 1/2$, and are calculated for various values of $E_J/E_C$. The zero point energy is chosen as the bottom of $m = 0$ level. For increasing values of $E_J/E_C$, $E_m$ becomes more robust against fluctuations in $n_g$ arising from environmental noise, whereas the anharmonicity ($E_{\delta} = E_{01} - E_{12}$) reduces. $E_J/E_C$ is chosen between 10 and 50 for transmon qubits in order to obtain robustness with sufficient anharmonicity. Under this regime, $E_{01} \approx \sqrt{8E_CE_J}$, and $E_{\delta} \approx E_C/2$.  (c) Schematic of a transmon qubit capacitively coupled to a superconducting resonator for initialization, readout and control \cite{koch2007pra}. The capacitance between various entities of the transmon-resonator system are also labeled. (d) Equivalent circuit of a transmon coupled to the resonator \cite{koch2007pra}. \href{https://journals.aps.org/pra/abstract/10.1103/PhysRevA.76.042319}{Figures \ref{fig:TransmonQubitPlatforms}(b), \ref{fig:TransmonQubitPlatforms}(c) and \ref{fig:TransmonQubitPlatforms}(d) are reprinted from Ref. \cite{koch2007pra} with permission from the American Physical Society (APS)}.}
\label{fig:TransmonQubitPlatforms}
\end{figure*}

The device geometry for \emph{transmon} qubits consists of two superconducting islands that are coupled through two Josephson junctions and a large capacitance between them (Fig.  \ref{fig:TransmonQubitPlatforms}a). The Josephson junctions act as non-linear inductors, and the device thereby constitutes an $LC$ oscillator circuit \cite{girvin2011circuit}. The resulting energy levels in the device depend on two quantities : (i) \emph{Charging energy} ($E_C$) of the superconducting island determined by the total capacitance ($C$) in parallel to the junctions, and (ii) \emph{Josephson energy} ($E_J$) determined by Josephson inductance $L$, which in turn is dependent on the critical current of the junctions. $E_C$ and $E_J$ are given by $e^2/2C$ and $\hbar^2/\left(4e^2L\right)$ respectively, where $e$ is the electron charge, and $\hbar$ is the reduced Planck constant. The ratio $E_J/E_C$ determines the qubit-type encoded by the Jospehson junctions, resulting in a variety of superconducting qubits such as transmon, flux or phase qubits \cite{wendin2017rpp}. 

\par
For the popular transmon qubits, $C$ is chosen such that $E_J/E_C$ lies between 10 and 50 \cite{koch2007pra}, and the \emph{charge states} on the superconducting island encode quantum information. The qubit splitting is given by $E_{01} \approx \sqrt{8E_CE_J}$, and is $\sim 5$ GHz in units of frequency (Fig.  \ref{fig:TransmonQubitPlatforms}b). The difference between the qubit splitting and the other splittings in the system is often called the anharmonicity. For large values of $E_J/E_C$, the anharmonicity $E_{\delta} \approx E_C/2$, and is of the order $\sim 100$ MHz in units of frequency. The above choice of $E_J/E_C$ between 10 and 50 allows for a robust qubit with sufficient anharmonicity.  As shown in Fig.  \ref{fig:TransmonQubitPlatforms}a, the dimensions of transmon qubits are few tens of $\mu$m, enabling large-scale solid-state fabrication with techniques adapted from the microelectronics industry.      

\par
To perform quantum operations, the transmon qubits are commonly placed adjacent to a \emph{superconducting resonator} (Fig.  \ref{fig:TransmonQubitPlatforms}c), and is capacitively coupled to it  (Fig.  \ref{fig:TransmonQubitPlatforms}d)  \cite{koch2007pra, blais2004pra}. Here, the qubit-resonator system is designed to be in the \emph{dispersive} regime, where the detuning ($\Delta \sim$ 100 MHz) between qubit and the photonic mode of the resonator is much larger than the coupling ($g \sim$ 10 MHz) between them. In this regime, the shift in the resonator transmission frequency from its fundamental mode frequency is given by $\pm g^2/\Delta$, where the sign (+ or -) depends on the qubit state \cite{blais2004pra}. By applying microwave pulses to the resonator, and measuring its transmission, the qubit state can hence be readout.

\par
Resonant \emph{microwave pulses} can be used to control the qubits, as the qubit splitting is $\sim 5$ GHz. Qubit control timescales are a few hundreds of nanoseconds depending on the quantum gate operation, and are much faster than that of trapped ion and silicon spin qubits. Measurement of the qubit, and its subsequent control also aids in deterministic initialization of the qubit state.

\par
Two qubits which are significantly detuned from the resonator, can be coupled to each other via the resonator. The coupling rate between the qubits is given by $\frac{g_1 g_2}{2}\left(1/\Delta_1 + 1/\Delta_2\right)$, where $g_1$ and $g_2$ are their individual coupling strengths to the resonator, $\Delta_1$ and $\Delta_2$ are their detunings to the resonator \cite{majer2007n}. However, the effective coupling rates ($\sim$ MHz) between the qubits, will still be smaller than the detunings ($\sim 300$ MHz) between them, caused by differences in the qubit splittings during manufacturing. As a result, the resonance frequency of each qubit will be determined by the state of the other qubit, similar to the electron/nuclear spin qubit splittings shown in Fig.  \ref{fig:SiliconQubitPlatforms}d. This enables conditional rotation of one qubit, dependent on the state of the other qubit, and hence a CNOT gate. Alternatively, direct \emph{capacitive coupling} between two adjacent transmon qubits can also be leveraged for demonstrating CNOT gates. However, using only direct capacitive coupling between the qubits leads to significant cross talk when they are incorporated in a large-scale architecture.

\par
Compared to silicon and trapped-ion qubits, the relaxation and coherence times of superconducting qubits are short. The main sources of decoherence arise from coupling of the qubits to additional two level systems present in the bulk/interfaces of the device, non-equilibrium quasi-particles generated from stray-infrared light, and radiation to additional modes present in device \cite{martinis2014arxiv,dial2016bulk}. The relaxation rate has also been shown to be exponentially dependent on the temperature, due to the qubit interaction with thermal photons \cite{koch2007pra}. As a result, extremely low temperatures $\sim 20$ mK are necessary for high-fidelity operation of qubits. Different device designs and operation regimes during the last decade have resulted in improvements in the relaxation and coherence times by several orders of magnitude. 
Dephasing times currently is of the order of $\sim 100$ $\mu$s. 

\par
The Josephson energy is strongly determined by the critical current across the junction, which in turn is dependent on the superconducting energy gap and the normal resistance ($R_n$) of the Josephson junction when it is operated above the critical temperature \cite{ambegaokar1963tunneling}. $R_n$ is determined by the thickness (few nm) of the Josephson junction, and can be variable across different devices. This results in \emph{non-uniform qubit splittings} across devices, with an in-homogeneity of $\sim$ 300 MHz. Another significant challenge is the \emph{large size} (several tens of $\mu$m) of superconducting qubits, limiting the number of qubits that can be coupled to each other via a single resonator, which spans about a centimeter. Scaling up current demonstrations to a large-scale architecture with millions of well-connected qubits operating at extremely low temperature will benefit strongly by a reduction in the size of the qubits \cite{bosman2017nqi}. 

\par
While a standalone tool for designing superconducting qubits is non-existent, parameters such as the capacitance (for determining $E_C$) and inductance (for determining $E_J$) can be estimated with classical electrostatic and electromagnetic packages such as FastCap and FastHenry respectively. Microwave software such as TXLINE (in AWR Microwave Office) has been used to design and estimate the characteristic impedance of the superconducting  resonator, that aids to readout, control and couple the qubits. In addition, the electromagnetic fields experienced by the superconducting qubits, can be obtained by solving the Maxwell's Equations with high-frequency electromagnetic simulators such as ANSYS-HFSS. As for silicon and trapped-ion qubits, the qubit dynamics can also be obtained by solving the simplified Hamiltonian with mathematical packages.

\section{Verification and Validation of Quantum Devices}
\label{sec:vv}
%

In spite of the great progress in fabrication and control of qubits, today's quantum computing devices are far noisier and error-prone than conventional digital circuits.
Bit error probabilities of $10^{-3}-10^{-2}$ per qubit per operation (or per clock cycle) are typical.
Even with continued progress in qubit technologies, it is unlikely that the errors incurred by physical qubits will ever become negligible.
Thus understanding and mitigating fault processes in qubit devices is a critical aspect of quantum computer development.
Correspondingly, the experimental testing of qubit devices primarily concerns the accuracy and reliability of hardware operation rather than the correctness of the circuit logic.
Experimental testing is essential here: since qubit devices are designed to perform computations that would otherwise be impractical or impossible, realistic simulation at the device scale is not a viable option.

Qubit device testing may be divided into two broad categories: characterization, wherein the goal is to obtain a detailed model of a device's fault modes; and benchmarking, wherein the goal is to determine a few high-level performance metrics.
Characterization is the more costly type of testing but can provide important insights leading to fault mitigation strategies or improved devices.
For simply assessing the performance of a device, benchmarking is more practical.

\paragraph{Benchmarking: Metrics and Techniques}
The most basic performance metric is the probability that the device outputs the correct state. In the context of quantum mechanics, this corresponds to the inner product (or ``overlap'') between the output state and the intended state, which is called the \emph{fidelity}. The infidelity, defined as 1 minus the fidelity, quantifies the amount of error in the output state. Another common way of quantifying the output error is in terms of the geometric distance between the output state and the target state in the complex vector space.
\par
If a qubit device is used to output a specific quantum state, e.g. some reference state or resource state, the fidelity of the output with respect to this known state can be estimated by measuring random subsets of qubits along various directions of the Bloch sphere \cite{Flammia2011, daSilva2011}.
In such cases, the experimental cost scales favorably with the register size.
However, a qubit device would be used to perform a wide variety of computations each with a different output state, and these output states presumably cannot be computed by any conventional means.
In this case one desires experimental metrics that  allow one to estimate or bound the fidelity of the device output for any computation it performs. 
The state-of-the-art approach for this purpose is Randomized Benchmarking \cite{Magesan2012-PRA} (RB).
RB is a technique for assessing how much, on average, each operation decreases the output fidelity.
Essentially, RB involves measuring the final fidelity of a qubit for random operation sequences of varying lengths.
For weak uncorrelated errors, the fidelity decays exponentially as a function of sequence length.
The RB decay constant is broadly interpreted as the average error per gate, an obviously useful performance metric.
Extensions of RB have been devised to yield operation-specific error metrics \cite{Magesan2012-PRL, Kimmel2014}, to incorporate multi-qubit operations \cite{Gaebler2012}, to include qubit loss \cite{Chasseur2015}, and to assess cross talk \cite{Gambetta2012}.
While RB remains a very popular benchmarking method, its underlying fault model is not universal; hence RB in its current form may not be enitrely valid or accurate as engineering efforts continue to make the simple fault modes assumed by RB less and less prominent \cite{Ball2016}.
Additionally, it has been noted that relating RB decay constants to operation fidelities is subtly problematic \cite{Proctor2017}.

\paragraph{Characterization via Quantum Tomography}
An alternative to benchmarking is to thoroughly characterize the fault modes of the device.
Since the output state of a quantum circuit is exponentially large in the number of qubits, characterization of a quantum circuit as a whole is generally infeasible.
The established strategy is to characterize each operation of a qubit device as completely as possible, so that the result of any given sequence of operations can (in principle) be predicted accurately.
The general name for this strategy is \emph{quantum tomography}, a name derived from the medical imaging technique in which a 3-dimensional image of a subject is reconstructed from a set of 2-dimensional projections.
In a similar manner, quantum tomography reconstructs a quantum state or operation from multiple measurements, each of which reveals a particular projection of the state. This reconstruction is based on the fact that a quantum state is uniquely specified by the probability distributions for certain characteristic quantities of a physical system. (For a spin qubit, the characteristic quantities are the projection of the spin along three independent spatial directions.)
State tomography is the determination of the quantum state via statistical estimation of these characteristic distributions.
Tomographic methods can also be used to characterize qubit operations.
A qubit operation can be thought of as a linear transformation of the characteristic probability distributions.
Quantum process tomography is the determination of the transformation matrix by characterizing the output state for each possible input state, or more precisely, for a set of linearly independent states that span the state space.
\par
Quantum tomography as just described requires well-calibrated measurements, whereas qubit measurements are among the device operations that need to be characterized. This problem is overcome with Gate Set Tomography \cite{Merkel2013, Blume-Kohout2017}, the state-of-the-art method for detailed characterization of qubit devices. Gate set tomography involves tomographic measurements of many different sequences of device operations.
These sequences, which range in length up to hundreds or thousands of operations, are carefully chosen to reveal all possible types of qubit errors.
The data is then fit to a highly nonlinear model using a sophisticated procedure, yielding a self-consistent model of all of a device's operations, including the measurement operations themselves.
Gate Set Tomography has been used to characterize and significantly improve the control of trapped ion qubits \cite{Dehollain2016}.

\paragraph{Other Approaches}
In addition to Randomized Benchmarking and Gate Set Tomography, a number of other testing approaches have been developed.  Some of these remain theoretical proposals, while others have had at least limited experimental demonstrations.

One approach is to test a quantum device utilizing another quantum device, either as a reference or as a resource to perform more powerful quantum-based tests  \cite{Macchiavello2013}. This line of approach stands to greatly reduce the cost of quantum device characterization, but it requires the availability of well-characterized quantum circuits that are similarly difficult to certify.

Another approach is to exploit prior knowledge to reduce the cost of conventional benchmarking and tomographic methods.
For example, adaptive testing based on Bayesian principles can significantly accelerate both randomized benchmarking \cite{Granade2015} and tomography \cite{Huszar2012, Mahler2013}.
In the case that the state or operation in question has some known characteristics (e.g. it has low rank or belongs to a certain symmetry class), specialized testing methods that are more efficient are applicable \cite{Chen2013, Toth2010}.
Related to this, the technique of compressive sensing has been adapted to the quantum domain and applied to the characterization of quantum states \cite{Gross2010}.

Other forms of testing may be categorized as model fitting, e.g. determining particular parameters of qubit dynamics, or assessing particular properties of the device output (e.g. purity or entanglement).
One recently-developed approach to characterizing the quality of many-qubit devices is to measure the distribution of output states produced by executing random quantum circuits \cite{Boixo2016}.
This reveals the extent to which the device can create and maintain superpositions of computational states, a key facet of the ``quantumness'' of quantum computation.

Finally, there is now a rapidly growing interest in the use of machine learning techniques for characterizing quantum systems.
Instead of attempting to match experimental data to an intrinsically quantum model that is likely to be intractable, researchers have begun to use neural nets to learn the behavior of quantum systems from experimental data \cite{Landon-Cardinal2012, Granade2012, Carleo2017,Deng2017}.
The learning process implicitly creates a tractable model of the quantum system.

\section{Quantum Circuit Design}
\label{sec:circuits}
%
A quantum circuit provides a formal representation of the register elements and sequences of gates required for the implementation of a quantum algorithm. As summarized in Sec.~\ref{sec:principles}, gates represent quantum mechanical operators that address one or more register elements. By design, these quantum-mechanical operators are reversible (Hermitian) and can be represented as unitary matrices \cite{Yanofsky2008quantumbook}. In this section, we review the design and testing of quantum circuits with an emphasis on arithmetic operations, such as addition, subtraction and multiplication, which are required in the  implementations of many quantum algorithms \cite{bhaskar} \cite{Yanofsky2008quantumbook}.  
\par
\begin{figure}[hbt]
\centering
\begin{subfigure}[hb]{2.5in}
\begin{center}
	\textsc{Clifford+T Gate Set}
\end{center}
\end{subfigure} \\ \begin{subfigure}[hb]{1in}
Hadamard Gate
\end{subfigure} \qquad \begin{subfigure}[hb]{.5in}
\[
 \Qcircuit @C=0.7em @R=0.5em @!R{
 & \gate{H} &  \qw & &   }
 \]
\end{subfigure} \qquad \begin{subfigure}[hb]{.75in}
\centering
  $ \frac{1}{\sqrt{2}}
  \begin{bmatrix}
    1 & 1  \\
    1 & -1 
  \end{bmatrix} $
\end{subfigure} \\ \begin{subfigure}[hb]{1in}

\end{subfigure} \qquad \begin{subfigure}[hb]{.5in}

\end{subfigure} \qquad \begin{subfigure}[hb]{.75in}

\end{subfigure} \\ \begin{subfigure}[hb]{1in}
T Gate
\end{subfigure} \qquad \begin{subfigure}[hb]{.5in}
\[
 \Qcircuit @C=0.7em @R=0.5em @!R{
 & \gate{T} &  \qw & &   }
 \]
\end{subfigure} \qquad \begin{subfigure}[hb]{.75in}
\centering
  $\begin{bmatrix}
    1 & 0  \\
    0 & e^{i \cdot \frac{\pi}{4}} 
  \end{bmatrix} $
\end{subfigure} \\ \begin{subfigure}[hb]{1in}

\end{subfigure} \qquad \begin{subfigure}[hb]{.5in}

\end{subfigure} \qquad \begin{subfigure}[hb]{.75in}

\end{subfigure} \\ \begin{subfigure}[hb]{1in}
\flushleft
Hermitian of T Gate
\end{subfigure} \qquad \begin{subfigure}[hb]{.5in}
\[
 \Qcircuit @C=0.7em @R=0.5em @!R{
 & \gate{T^{\dag}} &  \qw & &   }
 \]
\end{subfigure} \qquad \begin{subfigure}[hb]{.75in}
\centering
    $\begin{bmatrix}
    1 & 0  \\
    0 & e^{-i \cdot \frac{\pi}{4}} 
  \end{bmatrix} $
\end{subfigure} \\ \begin{subfigure}[hb]{1in}

\end{subfigure} \qquad \begin{subfigure}[hb]{.5in}

\end{subfigure} \qquad \begin{subfigure}[hb]{.75in}

\end{subfigure} \\ \begin{subfigure}[hb]{1in}
Phase Gate
\end{subfigure} \qquad \begin{subfigure}[hb]{.5in}
 \[
 \Qcircuit @C=0.7em @R=0.5em @!R{
 & \gate{S} &  \qw & &   }
 \]
\end{subfigure} \qquad \begin{subfigure}[hb]{.75in}
\centering
  $\begin{bmatrix}
    1 & 0  \\
    0 & i 
  \end{bmatrix} $
\end{subfigure} \\ \begin{subfigure}[hb]{1in}

\end{subfigure} \qquad \begin{subfigure}[hb]{.5in}

\end{subfigure} \qquad \begin{subfigure}[hb]{.75in}

\end{subfigure} \\ \begin{subfigure}[hb]{1in}
\flushleft
Hermitian of Phase Gate
\end{subfigure} \qquad \begin{subfigure}[hb]{.5in}
 \[
 \Qcircuit @C=0.7em @R=0.5em @!R{
 & \gate{S^{\dag}} &  \qw & &   }
 \]
\end{subfigure} \qquad \begin{subfigure}[hb]{.75in}
\centering
  $\begin{bmatrix}
    1 & 0  \\
    0 & -i 
  \end{bmatrix} $
\end{subfigure} \\ \begin{subfigure}[hb]{1in}

\end{subfigure} \qquad \begin{subfigure}[hb]{.5in}

\end{subfigure} \qquad \begin{subfigure}[hb]{.75in}

\end{subfigure} \\ \begin{subfigure}[hb]{1in}
Not Gate
\end{subfigure} \qquad \begin{subfigure}[hb]{.5in}
  \[
 \Qcircuit @C=0.7em @R=0.5em @!R{
 & \targ &  \qw & &   }
 \]
\end{subfigure} \qquad \begin{subfigure}[hb]{.75in}
\centering
 $\begin{bmatrix}
    0 & 1  \\
    1 & 0 
  \end{bmatrix}$
\end{subfigure} \\ \begin{subfigure}[hb]{1in}

\end{subfigure} \qquad \begin{subfigure}[hb]{.5in}

\end{subfigure} \qquad \begin{subfigure}[hb]{.75in}

\end{subfigure} \\ \begin{subfigure}[hb]{1in}
Feynman (CNOT) Gate
\end{subfigure} \qquad \begin{subfigure}[hb]{.5in}
  \[
 \Qcircuit @C=0.7em @R=0.5em @!R{
 & \ctrl{1} &  \qw & & \\ 
 & \targ &  \qw & & }
 \]
\end{subfigure} \qquad \begin{subfigure}[hb]{.75in}
\centering
  $\begin{bmatrix}
    1 & 0 & 0 & 0 \\
    0 & 1 & 0 & 0\\
    0 & 0 & 0 & 1 \\
    0 & 0 & 1 & 0 \\
  \end{bmatrix} $ 
\end{subfigure}
\caption{The  Clifford+T gate quantum gate set.}
\label{overview:CliddordT}
\end{figure}
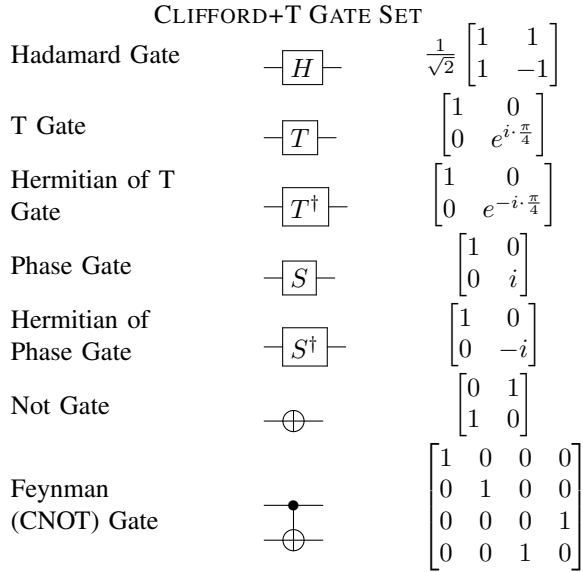

The design of quantum arithmetic circuits based on Clifford+T gates has caught the attention of researchers \cite{Lin} \cite{bhaskar} \cite{Boss2} \cite{edgard2017multiplier}. Figure \ref{overview:CliddordT} presents the quantum gates in the Clifford+T gate set with their matrix and graphic representations. 
The Clifford+T quantum gate set can be used to realize multi-qubit logic gates such as the Toffoli and Fredkin gates previously presented in the literature \cite{Toffoli1980toffoligate} \cite{Fredkin1982fredgate}. These multi-qubit gates will prove useful for describing the implementation of quantum circuits presented in this article.
\paragraph{CNOT Gate}

\begin{figure}[ht]
\centering
\begin{subfigure}[hb]{1in}
  \[
 \Qcircuit @C=0.7em @R=0.5em @!R{
 A & & \ctrl{1} &  \qw & & & A  & \\ 
 B & & \targ &  \qw & & & A \oplus B & }
 \]
\end{subfigure} \qquad \begin{subfigure}[hb]{.75in}
\centering
  $\begin{bmatrix}
    1 & 0 & 0 & 0 \\
    0 & 1 & 0 & 0\\
    0 & 0 & 0 & 1 \\
    0 & 0 & 1 & 0 \\
  \end{bmatrix} $ 
\end{subfigure}
\caption{The CNOT gate.   Matrix and graphic representations are shown.}
\label{overview:CNOTIntro}

\end{figure}
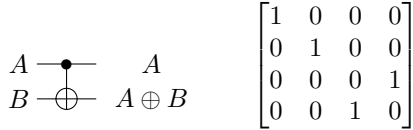

Figure \ref{overview:CNOTIntro} presents the matrix and graphic representations of the CNOT gate.  The CNOT gate is a Clifford+T gate (see figure \ref{overview:CliddordT}).  The CNOT gate is a 2 input, 2 output logic gate and has the mapping $A,B$ to $A,A \oplus B$. 

\paragraph{Toffoli Gate}

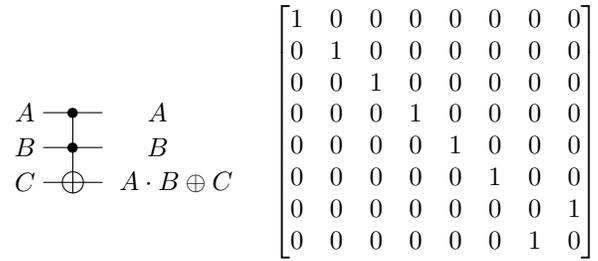
\begin{figure}[ht]
\centering
\begin{subfigure}[hb]{1.15in}
   \[
 \Qcircuit @C=0.7em @R=0.5em @!R{
A &  & \ctrl{1} &  \qw & & & A\\ 
B & & \ctrl{1} &  \qw & & & B\\ 
C & & \targ &  \qw & & & & A \cdot B \oplus C}
 \]
\end{subfigure} \qquad \begin{subfigure}[hb]{1.5in}
\centering
    $\begin{bmatrix}
    1 & 0 & 0 & 0 & 0 & 0 & 0 & 0\\
    0 & 1 & 0 & 0 & 0 & 0 & 0 & 0\\
    0 & 0 & 1 & 0 & 0 & 0 & 0 & 0\\
    0 & 0 & 0 & 1 & 0 & 0 & 0 & 0\\
    0 & 0 & 0 & 0 & 1 & 0 & 0 & 0\\
    0 & 0 & 0 & 0 & 0 & 1 & 0 & 0\\
    0 & 0 & 0 & 0 & 0 & 0 & 0 & 1 \\
    0 & 0 & 0 & 0 & 0 & 0 & 1 & 0
  \end{bmatrix} $
\end{subfigure}
\caption{The Toffoli gate.  Matrix and graphic representations are shown.}
\label{overview:ToffoliIntro}
\end{figure}

\begin{figure}[htb]
\flushleft
\small
\[
\Qcircuit @C=0.7em @R=0.5em @!R{
&	&\ctrl{2}		&\qw	&		&		&\qw &\gate{T}	&\qw		&\targ		&\qw			&\ctrl{2}		&\qw		&\ctrl{1}		&\gate{T^{\dag}}	&\qw			&\ctrl{2}		&\targ		&\qw		&\\
&	&\ctrl{1}		&\qw		& = 	&		&\qw		&\gate{T}	&\qw		&\ctrl{-1}		&\targ		&\qw			&\gate{T^{\dag}}	&\targ		&\gate{T^{\dag}}	&\targ		&\qw			&\ctrl{-1}		&\qw		&\\
&	&\targ		&\qw		&		&	&\gate{H}		&\gate{T}	&\qw	&\qw		&\ctrl{-1}		&\targ		&\qw	&\qw		&\gate{T}	&\ctrl{-1}		&\targ		&\gate{H}	&\qw		&
}
\]
\caption{The Toffoli gate and its Clifford+T quantum gate implementation \cite{Maslov}.}
\label{overview:clifTtoffoli}
\end{figure}
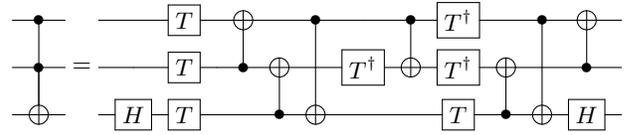
Figure \ref{overview:ToffoliIntro} presents the matrix and graphic representations of the Toffoli gate.  Figure \ref{overview:clifTtoffoli} shows an example Clifford+T quantum gate implementation of the Toffoli gate.  The Toffoli gate is a 3 input, 3 output logic gate that has the mapping $A,B,C$ to $A,B,A \cdot B \oplus C$.  With appropriate input combinations, the Toffoli gate can realize many logic operations such as AND, OR and NAND.  

\paragraph{Fredkin Gate}
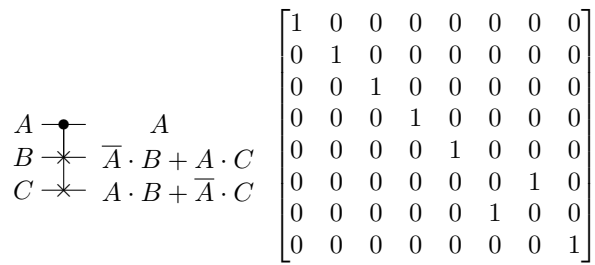
\begin{figure}[hb]
\flushleft
\begin{subfigure}[hb]{1.15in}
    \[
 \Qcircuit @C=0.7em @R=1.0em @!R{
A &  & \ctrl{1} &  \qw & & & & A\\ 
B & & \qswap \qwx &  \qw & & & & & \overline{A} \cdot B + A \cdot C\\ 
C & & \qswap \qwx &  \qw & & & & & A \cdot B + \overline{A} \cdot C}
 \]
\end{subfigure} \qquad \begin{subfigure}[hb]{1.5in}
\centering
    $\begin{bmatrix}
    1 & 0 & 0 & 0 & 0 & 0 & 0 & 0\\
    0 & 1 & 0 & 0 & 0 & 0 & 0 & 0\\
    0 & 0 & 1 & 0 & 0 & 0 & 0 & 0\\
    0 & 0 & 0 & 1 & 0 & 0 & 0 & 0\\
    0 & 0 & 0 & 0 & 1 & 0 & 0 & 0\\
    0 & 0 & 0 & 0 & 0 & 0 & 1 & 0\\
    0 & 0 & 0 & 0 & 0 & 1 & 0 & 0 \\
    0 & 0 & 0 & 0 & 0 & 0 & 0 & 1
  \end{bmatrix} $
\end{subfigure}
\caption{The Fredkin gate.   Matrix and graphic representations are shown.}
\label{overview:FredkinIntro}
\end{figure}

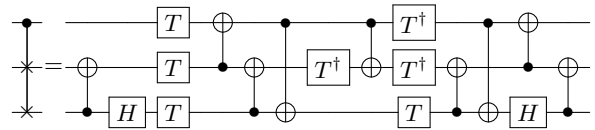
\begin{figure}[htb]
\flushleft
\small
\[
\Qcircuit @C=0.5em @R=0.4em @!R{
&	&\ctrl{2}		&\qw	&		&		&\qw &\qw &\gate{T}	&\qw		&\targ		&\qw			&\ctrl{2}		&\qw		&\ctrl{1}		&\gate{T^{\dag}}	&\qw			&\ctrl{2}		&\targ		&\qw		&\qw &\\
&	&\qswap \qwx		&\qw		& = 	&		&\targ &\qw		&\gate{T}	&\qw		&\ctrl{-1}		&\targ		&\qw			&\gate{T^{\dag}}	&\targ		&\gate{T^{\dag}}	&\targ		&\qw			&\ctrl{-1}		&\targ &\qw		&\\
&	&\qswap \qwx		&\qw		&		&	&\ctrl{-1} &\gate{H}		&\gate{T}	&\qw	&\qw		&\ctrl{-1}		&\targ		&\qw	&\qw		&\gate{T}	&\ctrl{-1}		&\targ		&\gate{H}	&\ctrl{-1} &\qw		&
}
\]
\caption{The Fredkin gate and Clifford+T quantum gate implementation \cite{Maslov}.}
\label{overview:clifTfedkin}
\end{figure}
Figure \ref{overview:FredkinIntro} presents the matrix and graphic representations of the Fredkin gate.  Figure \ref{overview:clifTfedkin} shows how the Fredkin gate can be realized using Clifford+T quantum gates.  The Fredkin gate is a 3 input, 3 output logic gate that has the mapping $A,B,C$ to $A,\overline{A} \cdot B + A \cdot C, A \cdot B + \overline{A} \cdot C $.  Like the Toffoli gate, the Fredkin gate can realize many logic operations.  
\par
Recent proposals for the realizations of reversible logic gates and quantum circuits have focused on the fault tolerant Clifford+T gate set due to its demonstrated tolerance to noise errors \cite{Maslov} \cite{Devitt}. The potential fault-tolerant implementations of these gates would play an important role in overcoming the noise observed in current quantum computing devices \cite{Maslov} \cite{Devitt} \cite{Hastings2017Tcost}. While fault-tolerant implementations can help to tolerate limited amounts of noise \cite{Mosca2} \cite{Knill2004faulttolerantQC}, it is important to note that the overhead associated with the implementation of such gates can be significant \cite{Devitt} \cite{Hastings2017Tcost}. Therefore, an important concern for optimal quantum circuit design is to account for the resource overheads associated with each gate. For example, there is an increased cost to realize the fault-tolerant T gate, making T-count and T-depth important performance measures for fault tolerant quantum circuit design \cite{Mosca2} \cite{Gosset}.  
\par
Resource evaluation of quantum circuits in terms of T-count and T-depth is of interest to researchers because of the high implementation costs of the T gate \cite{Hastings2017Tcost} \cite{Devitt}.  The number of qubits in a quantum circuit is a resource measure of interest because of the limited number of qubits available on existing quantum computers \cite{Lu2017computersaresmall} \cite{Song2017computersaresmall}.  We now define the T-count, T-depth, and qubit cost resource measures.
\begin{itemize}
\item Qubit cost: Qubit cost is the total number of qubits required to design the quantum circuit. 
\item T-count: T-count is the total number of T gates used in the quantum circuit
\item T-depth: T-depth is the number of T gate layers in the circuit, where a layer consists of quantum operations that can be performed simultaneously.
\end{itemize}


\par
Because quantum operators are reversible (Hermitian), in any quantum circuit there is a one-to-one mapping between  input and output vectors.  To maintain one-to-one mapping between  input and output vectors there is an associated overhead of ancillae and garbage outputs.  Any constant inputs used in the quantum circuit are called ancillae.  Garbage output refers to any output which exists in the quantum circuit to preserve one-to-one mapping but are neither one of the primary inputs nor a useful output.   The inputs regenerated at the circuit output are not considered garbage outputs \cite{Fredkin1982fredgate}. Ancillae and garbage outputs are circuit overheads that need to be minimized.  An ideal quantum circuit must be garbageless in nature.  Bennett's garbage removal scheme can be applied to design garbageless quantum circuits\cite{Bennett1973trashremoval}. 
\par
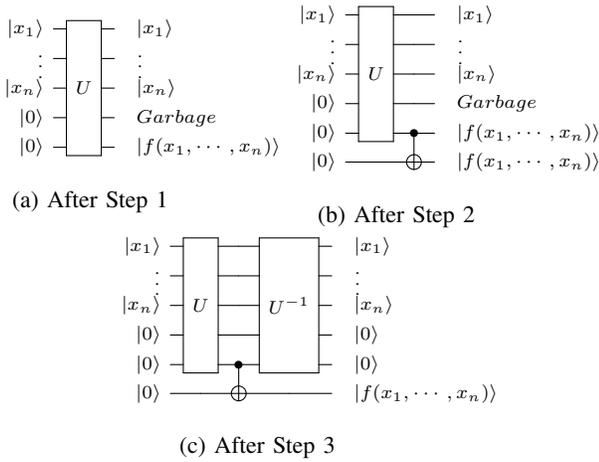
\begin{figure}[htb]
	\scriptsize
	\flushleft
\begin{subfigure}[thb]{1.25in}
		\flushleft
\[
\Qcircuit @C = .7em @R = .7em {
& \lstick{\ket{x_1}} & \multigate{4}{U} & \qw & \rstick{\ket{x_1}}\\
\vdots & & \ghost{U} & \qw & & \vdots \\
& \lstick{\ket{x_n}} & \ghost{U} & \qw & \rstick{\ket{x_n}}\\
& \lstick{\ket{0}} & \ghost{U} & \qw & \rstick{Garbage} \\
& \lstick{\ket{0}} & \ghost{U} & \qw & \rstick{\ket{f(x_1,\cdots,x_n) }} \\
}
\]
\caption{After Step 1}
\end{subfigure}  \qquad \begin{subfigure}[thb]{1.5in}

\[
\Qcircuit @C = .7em @R = .7em {
& \lstick{\ket{x_1}} & \multigate{4}{U} & \qw &\qw & \rstick{\ket{x_1}}\\
\vdots & & \ghost{U} & \qw & \qw & & \vdots \\
& \lstick{\ket{x_n}} & \ghost{U} & \qw & \qw & \rstick{\ket{x_n}}\\
& \lstick{\ket{0}} & \ghost{U} & \qw & \qw & \rstick{Garbage} \\
& \lstick{\ket{0}} & \ghost{U} & \ctrl{1}  & \qw & \rstick{\ket{f(x_1,\cdots,x_n) }} \\  
& \lstick{\ket{0}} & \qw &\targ& \qw & \rstick{\ket{f(x_1,\cdots,x_n) }} \\
}
\]
\caption{After Step 2}
\end{subfigure} \\ \begin{subfigure}[thb]{3in}

\[
\Qcircuit @C = .7em @R = .7em {
& \lstick{\ket{x_1}} & \multigate{4}{U} & \qw &\multigate{4}{U^{-1}} & \qw &\rstick{\ket{x_1}}\\
\vdots & & \ghost{U} & \qw & \ghost{U^{-1}} & \qw & & \vdots\\
& \lstick{\ket{x_n}} & \ghost{U} & \qw & \ghost{U^{-1}} & \qw & \rstick{\ket{x_n}}\\
& \lstick{\ket{0}} & \ghost{U} & \qw & \ghost{U^{-1}} & \qw & \rstick{\ket{0}} \\
& \lstick{\ket{0}} & \ghost{U} & \ctrl{1} & \ghost{U^{-1}} & \qw & \rstick{\ket{0}} \\  
& \lstick{\ket{0}} & \qw &\targ& \qw & \qw & \rstick{\ket{f(x_1,\cdots,x_n)}} \\
}
\]
\caption{After Step 3}
\end{subfigure}
\caption{Illustration of the Bennett's garbage removal scheme.}
\label{overview:bennet}
\end{figure}

For example, consider a quantum circuit that has a significant number of garbage outputs. The garbage outputs can be removed by using Bennett's garbage removal scheme \cite{Bennett1973trashremoval}.  Figure \ref{overview:bennet} illustrates the Bennett's garbage removal scheme.  Let $U$ represent an arbitrary quantum circuit that performs $f(x_1,x_2,\cdots,x_{n-1},x_n)$ and let $U^{-1}$ represent its logical reverse.  

Bennett's garbage removal scheme is a three-step progress. After $U$ is applied, all desired outputs are copied to ancillae with CNOT gates.  Then, $U^{-1}$ is applied to the qubits of the original circuit $U$.   Thus, at the end of computation, the garbage outputs have been restored to their initial values.  
\subsection{Quantum Arithmetic Circuits}
\label{overview:arithmetic}
The quantum logic gates presented in the previous section can be combined to create quantum arithmetic circuits to implement quantum algorithms.  Quantum circuits for arithmetic operations such as addition, subtraction and multiplication based on these gates have been proposed the literature \cite{Lin} \cite{Boss2} \cite{edgard2017multiplier} \cite{Thapliyal2016addsub}, and in this section, we illustrate a recent quantum addition circuit and a recent quantum multiplication circuit.

\paragraph{Quantum Circuit for Addition}
\label{overview:add}

We show an example of a quantum ripple carry addition circuit with no input carry presented in \cite{Boss2}.  
Consider the addition of two $n$-bit numbers $a$ and $b$ stored at quantum registers $\ket{A}$ and $\ket{B}$ respectively.  Further, let quantum register location $\ket{A_n}$ be initialized with $z = 0$.  At the end of computation, the quantum register $\ket{B}$ will have the values $s_{n-1:0}$ while the quantum register $\ket{A}$ keeps the value $a$.  The additional quantum register location $\ket{A_n}$ that initially stored the value $z$ will have the value $s_n$ at the end of computation.
Here $s_i$ is the sum bit and is defined as:

\begin{equation}
s_i = \begin{cases}
a_i \oplus b_i \oplus c_i & \qquad \text{if } 0 \leq i \leq n-1 \\
c_n & \qquad \text{if } i = n \\
\end{cases}
\label{overview-equation:1}
\end{equation}

Where $c_i$ is the carry bit and is defined as:

\begin{equation}
c_i = \begin{cases}
0 & \, \text{if } i = 0 \\
a_{i-1} \cdot b_{i-1} \oplus b_{i-1} \cdot c_{i-1} \oplus a_{i-1} \cdot c_{i-1} & \, \text{if } 1 \leq i \leq n \\   
\end{cases}
\label{md-equation:2}
\end{equation}

Figure \ref{overview:add} illustrates the complete addition circuit for the case of two 4 bit inputs $a$ and $b$.

\begin{figure}[htb]
\flushleft
\small
\[
\Qcircuit @C=0.3em @R=0.5em @!R{
\lstick{\ket{b_0}}&		&\qw			&\qw			&\qw			&\qw			&\ctrl{3}		&\qw			&\qw			&\qw			&\qw			&\qw			&\qw			&\qw			&\qw			&\qw			&\ctrl{3}		&\targ			&\qw			&\qw			&\qw			&\rstick{\ket{s_0}} \\
\lstick{\ket{a_0}}&		&\qw			&\qw			&\qw			&\qw			&\ctrl{2}		&\qw			&\qw			&\qw			&\qw			&\qw			&\qw			&\qw			&\qw			&\qw			&\ctrl{2}		&\ctrl{-1}		&\qw			&\qw			&\qw			&\rstick{\ket{a_0}} \\
\lstick{\ket{b_1}}&	&\targ			&\qw			&\qw			&\qw			&\qw			&\ctrl{3}		&\qw			&\qw			&\qw			&\qw			&\qw			&\qw			&\ctrl{3}		&\targ		&\qw			&\qw			&\qw				&\targ		&\qw				&\rstick{\ket{s_1}} \\
\lstick{\ket{a_1}}&	&\ctrl{-1}		&\qw			&\qw			&\ctrl{2}		&\targ		&\ctrl{2}		&\qw			&\qw			&\qw			&\qw			&\qw			&\qw			&\ctrl{2}		&\ctrl{-1}		&\targ				&\ctrl{2}		&\qw			&\ctrl{-1}		&\qw			&\rstick{\ket{a_1}} \\
\lstick{\ket{b_2}}&	&\targ		&\qw			&\qw			&\qw			&\qw			&\qw			&\ctrl{3}		&\qw			&\qw			&\qw			&\ctrl{3}		&\targ		&\qw			&\qw			&\qw			&\qw			&\qw						&\targ &\qw		&\rstick{\ket{s_2}} \\
\lstick{\ket{a_2}}&	&\ctrl{-1}			&\qw			&\ctrl{2}		&\targ		&\qw			&\targ		&\ctrl{2}		&\qw			&\qw			&\qw			&\ctrl{2}		&\ctrl{-1}		&\targ		&\qw			&\qw					&\targ		&\ctrl{2}				&\ctrl{-1}	&\qw	&\rstick{\ket{a_2}} \\
\lstick{\ket{b_3}}&	&\targ		&\qw			&\qw			&\qw			&\qw			&\qw			&\qw			&\ctrl{2}		&\qw			&\targ		&\qw			&\qw			&\qw			&\qw			&\qw			&\qw	&\qw	&\targ		&\qw 	&\rstick{\ket{s_3}} \\
\lstick{\ket{a_3}}&	&\ctrl{-1}		&\ctrl{1}		&\targ		&\qw			
&\qw			&\qw			&\targ		&\ctrl{1}		&\qw			
	&\ctrl{-1}		&\targ		&\qw			&\qw			
&\qw			&\qw			&\qw				&\targ			
&\ctrl{-1}	&\qw	&\rstick{\ket{a_3}} \\
\lstick{\ket{z}}&	&\qw			&\targ			&\qw			&\qw			
&\qw			&\qw			&\qw			&\targ		&\qw		&\qw			&\qw			&\qw			
&\qw			&\qw			&\qw			&\qw			
&\qw			&\qw			&\qw			&\rstick{\ket{z}} \\
}
\]

\caption{A 4-qubit example of the quantum ripple carry addition circuit with no input carry presented in \cite{Boss2}.}
\label{overview:add}

\end{figure}
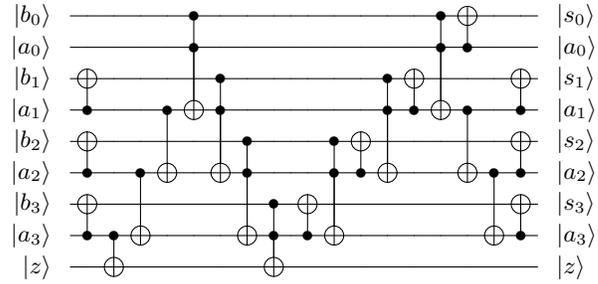

The carry bits $c_i$ are produced based on the inputs $a_{i-1},b_{i-1}$ and the carry bit $c_{i-1}$ from the previous stage.  Each generated carry bit $c_i$ is stored at the quantum register location $\ket{A_i}$ that initially stored the value $a_i$ for $0 \leq i \leq n-1$.  After the generated carry bits are used in further computation, each quantum register location $\ket{A_i}$ is restored to the value $a_i$ while each quantum register location $\ket{B_i}$ stores the sum bit $s_i$ for $0 \leq i \leq n-1$.  The restoration of $\ket{A_i}$ to the value $a_i$ eliminates all garbage outputs and transforming $\ket{B_i}$ to the sum $s_i$ cuts the ancillae cost to $1$.

\paragraph{Quantum Circuit for Multiplication}

We present an example of a quantum integer multiplication circuit that is presented in \cite{edgard2017multiplier}.   The quantum circuit is based on a novel design of a quantum conditional addition (\textit{Ctrl-Add}) circuit with no input carry and the Toffoli gate array.  The quantum multiplication circuit implements the shift and add multiplication algorithm.  As a result, the circuit will require a total of $n$ \textit{Ctrl-Add} circuits and Toffoli gate arrays.  The \textit{Ctrl-Add} circuits and Toffoli gate arrays are placed such that the shift operations are accomplished with no additional gates. 

 Consider the multiplication of two $n$ bit numbers $a$ and $b$ stored in quantum registers $\ket{A}$ and $\ket{B}$ respectively.  Further, consider a quantum register $\ket{P}$ of size $2 \cdot n + 1$ initialized to $z = 0$.  At the end of computation, the quantum registers $\ket{A}$ and $\ket{B}$ keep the values $a$ and $b$ respectively.  At the end of computation, the quantum register locations $\ket{P_{0:2\cdot n-1}}$ will have the product of $a$ and $b$.  The quantum register location $\ket{P_{2 \cdot n}}$ will be restored to the value $0$.

Figure \ref{overview:mult} illustrates the quantum integer multiplication circuit for the case of two 4 bit inputs $a$ and $b$.

\begin{figure}[htb]
\flushleft
\[
\Qcircuit @C=0.3em @R=0.5em @!R{
\lstick{\ket{B_0}}&	&\ctrl{5}			&\qw	&\qw &\qw &\qw &\qw &\qw &\qw	& \rstick{\ket{B_0}}\\
\lstick{\ket{B_1}}&	&\qw			&\qw	&\ctrl{4} &\qw	&\qw &\qw &\qw 
&\qw & \rstick{\ket{B_1}}\\
\lstick{\ket{B_2}}&	&\qw			&\qw	&\qw &\qw	&\ctrl{4} &\qw &\qw 
&\qw & \rstick{\ket{B_2}}\\
\lstick{\ket{B_3}}&	&\qw		&\qw	&\qw &\qw	&\qw &\qw &\ctrl{4} &\qw & 
\rstick{\ket{B_3}}\\
\lstick{\ket{A_{3:0}}}&	&\ctrl{1}		&\qw	&\ctrl{2} &\qw	&\ctrl{3} &\qw 
&\ctrl{4} &\qw & \rstick{\ket{A_{3:0}}} \\
\lstick{\ket{0}}&	&\multigate{3}{\begin{sideways} Toffolis \end{sideways}}			&\qw	&\qw &\qw	&\qw &\qw &\qw &\qw & \rstick{\ket{P_0}}\\
\lstick{\ket{0}}&	&\ghost{\begin{sideways} Toffolis 
\end{sideways}}			&\qw	&\multigate{5}{ \begin{sideways} Ctrl-Add 
\end{sideways}} &\qw	&\qw &\qw &\qw &\qw &\rstick{\ket{P_1}} \\
\lstick{\ket{0}}&	&\ghost{\begin{sideways} Toffolis 
\end{sideways}}			&\qw	&\ghost{ \begin{sideways} Ctrl-Add 
\end{sideways}} &\qw	&\multigate{5}{ \begin{sideways} Ctrl-Add 
\end{sideways}}  &\qw &\qw &\qw &\rstick{\ket{P_2}} \\
\lstick{\ket{0}}&	&\ghost{\begin{sideways} Toffolis 
\end{sideways}}			&\qw	&\ghost{ \begin{sideways} Ctrl-Add 
\end{sideways}} &\qw &\ghost{ \begin{sideways} Ctrl-Add \end{sideways}}  
&\qw 	&\multigate{5}{ \begin{sideways} Ctrl-Add \end{sideways}}  &\qw 
&\rstick{\ket{P_3}}\\
\lstick{\ket{0}}&	&\qw			&\qw	&\ghost{ \begin{sideways} Ctrl-Add \end{sideways}} &\qw	&\ghost{ \begin{sideways} Ctrl-Add \end{sideways}} &\qw &\ghost{\begin{sideways} Toffolis \end{sideways}} &\qw &\rstick{\ket{P_4}}\\
\lstick{\ket{0}}&	&\qw			&\qw	&\ghost{ \begin{sideways} Ctrl-Add \end{sideways}} &\qw	&\ghost{ \begin{sideways} Ctrl-Add \end{sideways}} &\qw &\ghost{\begin{sideways} Toffolis \end{sideways}} &\qw &\rstick{\ket{P_5}}\\
\lstick{\ket{0}}&	&\qw			&\qw	&\ghost{ \begin{sideways} Ctrl-Add 
\end{sideways}} &\qw	&\ghost{ \begin{sideways} Ctrl-Add \end{sideways}} 
&\qw &\ghost{\begin{sideways} Toffolis \end{sideways}} &\qw 
&\rstick{\ket{P_6}}\\
\lstick{\ket{0}}&	&\qw		&\qw	&\qw &\qw	&\ghost{ \begin{sideways} 
Ctrl-Add \end{sideways}} &\qw &\ghost{\begin{sideways} Ctrl-Add 
\end{sideways}} &\qw & \rstick{\ket{P_7}} \\
\lstick{\ket{0}}&	&\qw		&\qw	&\qw &\qw	&\qw &\qw &\ghost{ 
\begin{sideways} Ctrl-Add 
\end{sideways}} &\qw & 
\rstick{\ket{0}}
}
\]

\caption{A 4-qubit example of the quantum integer multiplication circuit presented in \cite{edgard2017multiplier}}
\label{overview:mult}

\end{figure}
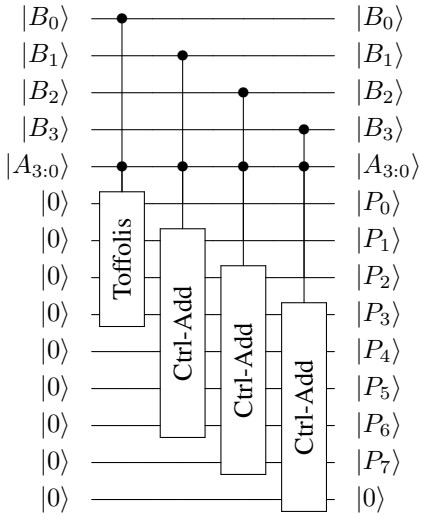

\begin{figure}[h!tb]
\flushleft
\footnotesize
\[
\Qcircuit @C=0.3em @R=0.5em @!R{
\lstick{\ket{ctrl}}&	&\qw			&\ctrl{9}		&\qw			&\qw			&\qw			&\qw			&\qw			&\qw			&\ctrl{10}		&\qw			&\ctrl{7}		&\qw			&\ctrl{5}		&\qw			&\ctrl{3}		&\qw			&\ctrl{1}			&\qw			&\qw			&\qw			&\rstick{\ket{ctrl}}  \\
\lstick{\ket{b_0}}&		&\qw			&\qw			&\qw			&\qw			&\ctrl{3}		&\qw			&\qw			&\qw			&\qw			&\qw			&\qw			&\qw			&\qw			&\qw			&\qw			&\ctrl{3}		&\targ			&\qw			&\qw			&\qw			&\rstick{\ket{s_0}} \\
\lstick{\ket{a_0}}&		&\qw			&\qw			&\qw			&\qw			&\ctrl{2}		&\qw			&\qw			&\qw			&\qw			&\qw			&\qw			&\qw			&\qw			&\qw			&\qw			&\ctrl{2}		&\ctrl{-1}		&\qw			&\qw			&\qw			&\rstick{\ket{a_0}} \\
\lstick{\ket{b_1}}&	&\targ			&\qw			&\qw			&\qw			&\qw			&\ctrl{3}		&\qw			&\qw			&\qw			&\qw			&\qw			&\qw			&\qw			&\ctrl{3}		&\targ		&\qw			&\qw			&\qw				&\targ		&\qw				&\rstick{\ket{s_1}} \\
\lstick{\ket{a_1}}&	&\ctrl{-1}		&\qw			&\qw			&\ctrl{2}		&\targ		&\ctrl{2}		&\qw			&\qw			&\qw			&\qw			&\qw			&\qw			&\qw			&\ctrl{2}		&\ctrl{-1}		&\targ				&\ctrl{2}		&\qw			&\ctrl{-1}		&\qw			&\rstick{\ket{a_1}} \\
\lstick{\ket{b_2}}&	&\targ		&\qw			&\qw			&\qw			&\qw			&\qw			&\ctrl{3}		&\qw			&\qw			&\qw			&\qw			&\ctrl{3}		&\targ		&\qw			&\qw			&\qw			&\qw			&\qw						&\targ &\qw		&\rstick{\ket{s_2}} \\
\lstick{\ket{a_2}}&	&\ctrl{-1}			&\qw			&\ctrl{2}		&\targ		&\qw			&\targ		&\ctrl{2}		&\qw			&\qw			&\qw			&\qw			&\ctrl{2}		&\ctrl{-1}		&\targ		&\qw			&\qw					&\targ		&\ctrl{2}				&\ctrl{-1}	&\qw	&\rstick{\ket{a_2}} \\
\lstick{\ket{b_3}}&	&\targ		&\qw			&\qw			&\qw			&\qw			&\qw			&\qw			&\ctrl{2}		&\qw			&\ctrl{2}		&\targ		&\qw			&\qw			&\qw			&\qw			&\qw			&\qw			&\qw		&\targ		&\qw 	&\rstick{\ket{s_3}} \\
\lstick{\ket{a_3}}&	&\ctrl{-1}		&\ctrl{1}		&\targ		&\qw			
&\qw			&\qw			&\targ		&\ctrl{2}		&\qw			
&\ctrl{2}		&\ctrl{-1}		&\targ		&\qw			&\qw			
&\qw			&\qw			&\qw				&\targ			
&\ctrl{-1}	&\qw	&\rstick{\ket{a_3}} \\
\lstick{\ket{z}}&	&\qw			&\targ			&\qw			&\qw			
&\qw			&\qw			&\qw			&\qw		&\targ		
&\qw		&\qw			&\qw			&\qw			
&\qw			&\qw			&\qw			&\qw			
&\qw			&\qw			&\qw			&\rstick{\ket{s_4}} \\
\lstick{\ket{z}}&	&\qw			&\qw		&\qw			&\qw			
&\qw			&\qw			&\qw			&\targ			
&\ctrl{-1} 		&\targ			&\qw			&\qw			
&\qw			&\qw			&\qw			&\qw			
&\qw			&\qw			&\qw			&\qw			&\rstick{\ket{z}}
}
\]
\caption{A 4-qubit example of the quantum \textit{Ctrl-Add} circuit with no input carry presented in \cite{edgard2017multiplier}.}
\label{overview:ctrladd}
\end{figure}
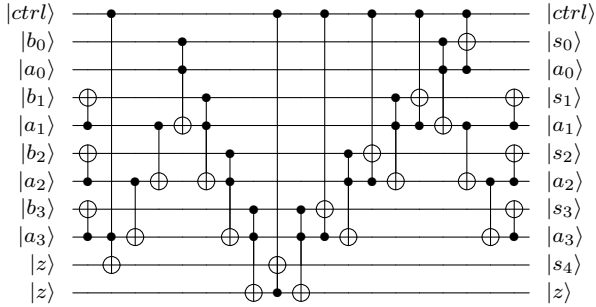

Figure \ref{overview:ctrladd} illustrates the quantum \textit{Ctrl-Add} circuit used in the quantum multiplier for the case of two 4 bit inputs $a$ and $b$.  The operation of the quantum \textit{Ctrl-Add} circuit is conditioned on the value of the qubit $\ket{ctrl}$.  When $\ket{ctrl = 1}$, the circuit performs addition.  The sum of $a$ and $b$ will appear on the quantum register that originally holds the value $b$ at the end of computation.  The quantum register that originally holds the value $a$ will be restored to the value $a$.  When $\ket{ctrl = 0}$, The quantum registers that initially hold the values $a$ and $b$ will be restored to the values $a$ and $b$ at the end of computation.

\paragraph{Application of Quantum Arithmetic Circuits in Taylor Series}
\label{overview:application}

In this section, we present an application of the quantum multiplication and quantum addition circuits presented in the previous section.  For this example, we consider a quantum circuit implementation of the Taylor series expansion.  Taylor series polynomials are used to approximate functions frequently used in scientific computing applications such as $sin(x)$, $ln(x)$ and $e^x$.  The value of a given function $f(x)$ near a point $c$ is estimated by computing the Taylor series equation shown below:

\begin{equation}
f(x) \approx \sum_{i = 0}^\infty \frac{f^i(c)}{i!} \cdot (x-c)^i
\label{overview:equation3}
\end{equation}

For this example, we only calculate the first three terms of the Taylor series.  What our circuit calculates is shown below:

\begin{equation}
f(x) \approx f(c) + f'(c) \cdot (x-c) + \frac{f''(c)}{2} \cdot (x-c)^2
\label{overview:equation4}
\end{equation}

 Consider the computation of the Taylor series for $f(x)$ centered at value $c$.  Let $c$ and $x$ be $n$ bit binary values stored in quantum registers $\ket{x}$ and $\ket{c}$, respectively.  Further, let $f(c)$, $f'(c)$ and $\frac{f''(c)}{2}$ be represented as $n$ bit binary numbers stored at quantum registers $\ket{f(c)}$, $\ket{f'(c)}$ and $\ket{\frac{f''(c)}{2}}$ respectively.  Lastly, consider quantum registers $\ket{Y_1}$, $\ket{Y_2}$, $\ket{Y_3}$ and $\ket{Y_4}$ that contain ancillae set to $0$.  At the end of computation, quantum register $\ket{Y_4}$ will have the first three terms of the Taylor series expansion.  The quantum registers $\ket{c}$, $\ket{x}$, $\ket{f(c)}$, $\ket{f'(c)}$ and $\ket{\frac{f''(c)}{2}}$ will be restored to the values $c$, $x$, $f(c)$, $f'(c)$ and $\frac{f''(c)}{2}$ at the end of computation.  Lastly, the quantum registers $\ket{Y_1}$, $\ket{Y_2}$ and $\ket{Y_3}$ that initially held ancillae will be restored to their initial values.

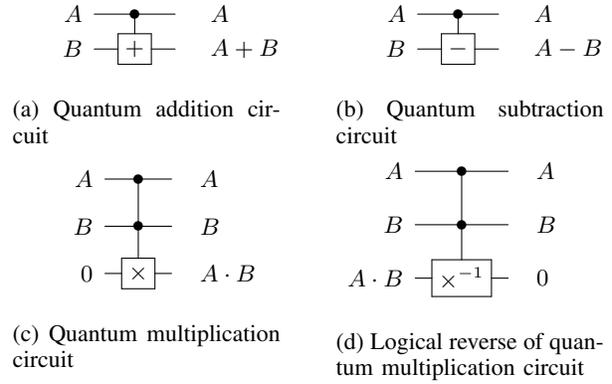
\begin{figure}
	\small
	\flushleft
	\begin{subfigure}[thb]{1.4in}
	\flushleft
		\[
		\Qcircuit @C=1em @R=.7em {
		\lstick{A} & \ctrl{1} & \qw & \rstick{A}\\
		\lstick{B} & \gate{+} & \qw & \rstick{A + B}\\
		} 
		\]
	\caption{Quantum addition circuit}
	\end{subfigure}  \qquad \begin{subfigure}[thb]{1.4in}
	\flushleft
		\[
		\Qcircuit @C=1em @R=.7em {
		\lstick{A} & \ctrl{1} & \qw & \rstick{A}\\
		\lstick{B} & \gate{-} & \qw & \rstick{A - B}\\
		} 
		\]	
	\caption{Quantum subtraction circuit}	
	\end{subfigure} \\
	\begin{subfigure}[thb]{1.4in}
		\flushleft
			\[
			\Qcircuit @C = .7em @R = .7em @!R{
			\lstick{A} & \ctrl{1} & \qw & \rstick{A}\\
			\lstick{B} & \ctrl{1} & \qw & \rstick{B}\\
			\lstick{0} & \gate{\times} & \qw & \rstick{A \cdot B} \\
			}
			\]
		\caption{Quantum multiplication circuit}
		\end{subfigure}  \qquad \begin{subfigure}[thb]{1.4in}
		\flushleft
			\[
			\Qcircuit @C = .7em @R = .7em @!R{
			\lstick{A} & \ctrl{1} & \qw & \rstick{A}\\
			\lstick{B} & \ctrl{1} & \qw & \rstick{B}\\
			\lstick{A \cdot B} & \gate{\times^{-1}} & \qw & \rstick{0} \\
			}
			\]
		\caption{Logical reverse of quantum multiplication 
		circuit}	
		\end{subfigure}
	\caption{Graphical representation of components used in the quantum Taylor series circuit in this article.}
\label{overview:Taylorpeices}
\end{figure}

The quantum Taylor series circuit is built from the quantum addition circuit, the quantum subtraction circuit, the quantum multiplication circuit and the logical reverse of the quantum multiplication circuit. Figure \ref{overview:Taylorpeices} shows the graphical representation of components used in the Taylor series circuit.  We will use a quantum subtraction circuit based on the ripple carry adder presented in this article that was presented in \cite{Thapliyal2016addsub}.     

\begin{figure}[htb]
  \[
\Qcircuit @C = .7em @R = .7em {
\lstick{\ket{b_0}} & \targ &\qw &\multigate{7}{\begin{sideways} Quantum Adder \end{sideways}} &\qw & \targ & \qw &\rstick{\ket{s_0}}\\
\lstick{\ket{a_0}} & \qw &\qw &\ghost{\begin{sideways} Quantum Adder \end{sideways}} &\qw & \qw & \qw &\rstick{\ket{a_0}}\\
\lstick{\ket{b_1}} & \targ &\qw &\ghost{\begin{sideways} Quantum Adder \end{sideways}} &\qw & \targ & \qw &\rstick{\ket{s_1}}\\
\lstick{\ket{a_1}} & \qw &\qw &\ghost{\begin{sideways} Quantum Adder \end{sideways}} &\qw & \qw & \qw &\rstick{\ket{a_1}}\\
\lstick{\ket{b_2}} & \targ &\qw &\ghost{\begin{sideways} Quantum Adder \end{sideways}} &\qw & \targ & \qw &\rstick{\ket{s_2}}\\
\lstick{\ket{a_2}} & \qw &\qw &\ghost{\begin{sideways} Quantum Adder \end{sideways}} &\qw & \qw & \qw &\rstick{\ket{a_2}}\\
\lstick{\ket{b_3}} & \targ &\qw &\ghost{\begin{sideways} Quantum Adder \end{sideways}} &\qw & \targ & \qw &\rstick{\ket{s_3}}\\
\lstick{\ket{a_3}} & \qw &\qw&\ghost{\begin{sideways} Quantum Adder \end{sideways}} &\qw & \qw & \qw &\rstick{\ket{a_3}}\\
}
\]
\caption{A 4-qubit example of the conversion of a quantum addition circuit into a subtraction circuit via the procedure in \cite{Thapliyal2016addsub}.}
\label{overview:subunit}
\end{figure}
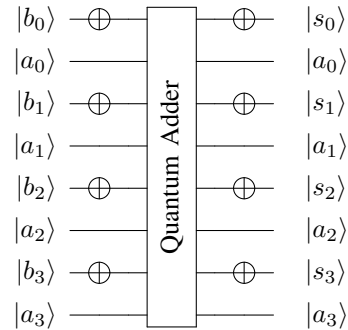

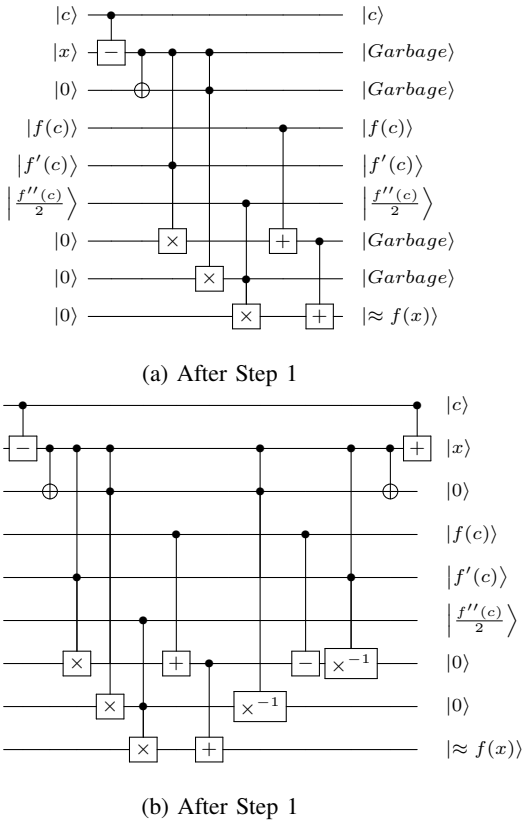
\begin{figure}[htb]
\centering
\begin{subfigure}[hb]{3in}
\scriptsize
\[
\Qcircuit @C=.5em @R=.7em @!R{
\lstick{\ket{c}} & \ctrl{1} & \qw & \qw & \qw & \qw & \qw & \qw & \qw & \rstick{\ket{c}}\\
\lstick{\ket{x}} & \gate{-} & \ctrl{1} & \ctrl{5} & \ctrl{5} & \qw & \qw & \qw & \qw & \rstick{\ket{Garbage}}\\
\lstick{\ket{0}} & \qw & \targ & \qw & \ctrl{5} & \qw & \qw & \qw & \qw & \rstick{\ket{Garbage}}\\
\lstick{\ket{f(c)}} & \qw & \qw & \qw & \qw & \qw & \ctrl{3} & \qw & \qw & \rstick{\ket{f(c)}} \\
\lstick{\ket{f'(c)}} & \qw & \qw & \ctrl{2} & \qw & \qw & \qw & \qw & \qw & \rstick{\ket{f'(c)}} \\
\lstick{\ket{\frac{f''(c)}{2}}} & \qw & \qw & \qw & \qw & \ctrl{3} & \qw & \qw & \qw & \rstick{\ket{\frac{f''(c)}{2}}} \\
\lstick{\ket{0}} & \qw & \qw & \gate{\times} & \qw & \qw & \gate{+} &\ctrl{2} &\qw & \rstick{\ket{Garbage}} \\
\lstick{\ket{0}} & \qw & \qw & \qw & \gate{\times} & \ctrl{1} & \qw & \qw & \qw & \rstick{\ket{Garbage}} \\
\lstick{\ket{0}} & \qw & \qw & \qw & \qw & \gate{\times} & \qw & \gate{+} & \qw & \rstick{\ket{\approx f(x)}} \\
} 
\]
\caption{After Step 1}
\end{subfigure} \\ \begin{subfigure}[hb]{3in}
\scriptsize
\[
\Qcircuit @C=.3em @R=0.7em @!R{
\lstick{\ket{c}} & \ctrl{1} & \qw & \qw & \qw & \qw & \qw & \qw & \qw & \qw &\qw & \qw & \qw & \ctrl{1} &\rstick{\ket{c}}\\
\lstick{\ket{x}} & \gate{-} & \ctrl{1} & \ctrl{5} & \ctrl{5} & \qw & \qw & \qw & \qw & \ctrl{3} &\qw & \ctrl{5} &\ctrl{1} & \gate{+} &\rstick{\ket{x}}\\
\lstick{\ket{0}} & \qw & \targ & \qw & \ctrl{5} & \qw & \qw & \qw & \qw & \ctrl{5} &\qw &\qw &\targ &\qw &\rstick{\ket{0}}\\
\lstick{\ket{f(c)}} & \qw & \qw & \qw & \qw & \qw & \ctrl{3} & \qw & \qw & \qw &\ctrl{3} &\qw &\qw &\qw &\rstick{\ket{f(c)}} \\
\lstick{\ket{f'(c)}} & \qw & \qw & \ctrl{2} & \qw & \qw & \qw & \qw & \qw & \qw & \qw &\ctrl{2} &\qw &\qw &\rstick{\ket{f'(c)}} \\
\lstick{\ket{\frac{f''(c)}{2}}} & \qw & \qw & \qw & \qw & \ctrl{3} & \qw & \qw & \qw & \qw &\qw &\qw &\qw &\qw &\rstick{\ket{\frac{f''(c)}{2}}} \\
\lstick{\ket{0}} & \qw & \qw & \gate{\times} & \qw & \qw & \gate{+} &\ctrl{2} \qw & \qw &\qw & \gate{-} & \gate{\times^{-1}} &\qw &\qw &\rstick{\ket{0}} \\
\lstick{\ket{0}} & \qw & \qw & \qw & \gate{\times} & \ctrl{1} & \qw & \qw & \qw & \gate{\times^{-1}} &\qw &\qw &\qw &\qw &\rstick{\ket{0}} \\
\lstick{\ket{0}} & \qw & \qw & \qw & \qw & \gate{\times} & \qw & \gate{+} & \qw & \qw &\qw &\qw &\qw &\qw &\rstick{\ket{\approx f(x)}} \\
} 
\]
\caption{After Step 1}
\end{subfigure}
\caption{Generation of the quantum circuit for the calculation of the first three terms of the Taylor series of $f(x)$: Steps 1-2}
\label{overview:Taylorcircuit}
\end{figure}

Figure \ref{overview:subunit} illustrates an example of the quantum subtraction circuit based on the design in \cite{Thapliyal2016addsub}.  The quantum circuit shown calculates  $\overline{\overline{b} + a}$ where $\overline{\overline{b} + a} = b-a $.  The circuitry used to calculate the sum bit $s_n$ is removed from the quantum adder because the circuitry is not needed to calculate $\overline{\overline{b} + a}$.  The steps to design the quantum Taylor series circuit are explained below.  Figure \ref{overview:Taylorcircuit} illustrates Steps 1 and 2.    

\begin{itemize}

\item Step 1: Calculate $f(x) \approx f(c) + f'(c) \cdot (x-c) + \frac{f''(c)}{2} \cdot (x-c)^2$.  We use the quantum multiplication circuit, quantum addition circuit and quantum subtraction circuit in this Step.  The result of the quantum subtraction circuit $x - c$ is copied to ancillae using an array of $n$ CNOT gates.

\item Step 2: Remove garbage output.  At the end of Step 1, three quantum registers ($\ket{Y_1}$, $\ket{Y_2}$ and $\ket{Y_3}$) that initially held ancillae are transformed to $f(c) + f'(c) \cdot (x-c)$, $(x-c)^2$ and $(x-c)$.  Further, at the end of computation, quantum register $\ket{x}$ that initially held the value $x$ has been transformed to the value $x-c$.  These outputs are  garbage outputs.  We use the logical reverse of the quantum multiplication circuit, the quantum adder, the quantum subtraction circuit and an array of CNOT gates to remove these garbage outputs.         

\end{itemize}

\section{Summary and Outlook}
\label{sec:conclusion}
%
We have summarized the basic features and requirements for quantum computing devices. This includes the fundamental criteria that a quantum computing device must implement as well as the the principles of operation for performing computation within the circuit model. We have reviewed the state of the art in three specific technologies currently being developed for quantum computing devices. Silicon spins, trapped ions, and superconducting transmons represent three of the leading approaches for quantum computing but these devices are still face fundamental research challenges. Therefore, methods to accurately characterize and benchmark the behavior of quantum computing devices plays an important role in design and testing. We have emphasized the necessity of statistical analysis to infer the operation of quantum devices. We have also discussed the design of optimal quantum circuits for the case of arithmetic operations, which represent an important use case for future quantum computing devices. These circuits were designed to minimize the occurrence of a specific instruction, the T gate, due to the expected complexity of fault-tolerant implementation. Such optimizations are expected to play a critical role in future device operation as trade-offs in gate and device complexity become more sophisticated.
\par
The design and testing of early quantum computing devices faces many near-term challenges. We have emphasized a small subset of the technologies currently under investigation for developing quantum computing devices. However, there are many more approaches to be considered, each with their own nuanced physics. This suggests that variations in the physics of each quantum computing technology may lead to different implementations for design and testing. Comparison across technologies will require standard calibration techniques that have yet to be developed. In addition, methods for quantifying well-defined metrics will be important for evaluating device performance. Current testing is focused on meeting the minimal criteria for functionality in the regime of noisy, error-prone, and faulty devices. Finally, we note that the current state of quantum computing remains focused on relatively small scale devices. Future devices, or networks of devices, are likely to include quantum registers with millions of elements and sequences with millions of highly parallelized instructions. Those devices and circuits will require more sophisticated methods for design and testing. 

\section*{Acknowledgments}
We thank Dr.~Yeun-Ming Shue for valuable insights into superconducting fabrication tools and Dr.~Raphael Pooser for guidance on quantum device development. This material is based upon work supported by the U.S. Department of Energy, Office of Science, Office of Advanced Scientific Computing Research, and Oak Ridge National Laboratory Directed Research and Development.

\end{document}